\documentclass[11pt,A4paper]{article}
\usepackage{jcappub}
\usepackage[utf8]{inputenc}
\usepackage{listings}
\usepackage{xcolor}
\usepackage{amsmath}
\usepackage{graphicx}
\usepackage{subfigure}
\usepackage[normalem]{ulem}

\title{Inflation in a  Gaussian Random Landscape}
\author{Lerh Feng Low, Richard Easther, Shaun Hotchkiss}
\date{June 2020}

\abstract{
Random, multifield functions can  set generic expectations for landscape-style cosmologies. We consider the inflationary implications of a landscape defined by a Gaussian random function, which is perhaps the simplest such scenario. Many key properties of this landscape, including the distribution of saddles as a function of height in the potential, depend only on its dimensionality, $N$, and a single parameter, $\gamma$, which is set by the power spectrum of the random function. We show that for saddles with a single downhill direction the negative mass term grows smaller, relative to the average mass, as $N$ increases, a result with potential implications for the $\eta$-problem in landscape scenarios.   For some power spectra Planck-scale saddles have $\eta \sim 1$ and eternal, topological inflation would  be common in these scenarios. Lower-lying saddles typically have large $\eta$, but the fraction of these saddles which would support inflation is computable, allowing us to identify which scenarios can deliver a universe that resembles ours. Finally, by drawing inferences about the  relative viability of different multiverse proposals we also illustrate ways in which quantitative analyses of multiverse scenarios are  feasible.
}

\begin{document}

\maketitle
\section{Introduction}
Some of the most fundamental questions in physics and cosmology come together in discussions of the hypothetical string landscape. This complex and difficult-to-specify function couples the many  scalar degrees of freedom found in flux-compactified string theory.  It can, in principle, contain a vast collection of minima each of which represents a potentially unique configuration of ``low  energy'' (i.e. sub-string scale)  physics in which the vacuum energy or cosmological constant is fixed by the value of the landscape potential at the minimum \cite{Douglas2019}. The landscape can be ``populated'' by a wide variety of cosmological mechanisms, including classical rolling, stochastic field evolution, quantum tunneling or classical transitions \cite{Coleman1980,Vilenkin1983, Starobinsky1988, Sasaki1988, Nakao1988, Kandrup1989,Easther:2009ft, Giblin:2010bd}.

The possible existence and properties of the landscape can be explored by direct analyses of  relevant stringy constructions and via more general consistency considerations such as those leading to  the swampland hypotheses \cite{Obied2018,Agrawal2018} and  weak gravity conjecture \cite{Arkani2007}.  However, many properties of the landscape are portrayed as a consequence of its complexity, rather than its detailed  form. Consequently, a further option is to use multidimensional random functions as proxies for the landscape itself. This strategy has been pursued in a variety of ways \cite{Aazami2006,Chen2012,Battefeld2012,Greene2013,Easther2016,Masoumi2016} but perhaps the most  conceptually straightforward is to directly investigate the properties of extrema of \emph{random functions} in ${\cal{O}}(100)$ dimensions  \cite{Henry2009,Frazer2011,Marsh2013,Dias2018,Bjorkmo2018,Yamada2018}, after noting that these points will determine both the likely range of vacuum energies and inflationary trajectories.

In Ref.~\cite{Low2020} we considered the distribution and properties of minima in a landscape modeled by an $N$-dimensional Gaussian random function.\footnote{The term ``Gaussian random field'' is common in the mathematical literature; we refer to random functions to avoid confusion with the $N$ scalar fields which are the independent variables of the landscape potential.}  For a number of interesting scenarios,  $P(\Lambda>0)$, where $\Lambda$ is the vacuum energy, can be vanishingly small, to the extent that in a landscape with $10^{500}$ minima  it is highly unlikely that even one of them has a positive vacuum energy.  
This paper examines the inflationary mechanisms that can operate in Gaussian random landscapes. The range of inflationary dynamics supported by a landscape will be reflected in the distributions of the eigenvalues of the Hessians. In the vicinity of an extremum these eigenvalues are the squares of the masses of the $N$ fields. One of the most interesting scenarios will be saddles with a single downhill direction, which we call `1-saddles'. As we will show, the expected magnitude of the downhill eigenvalue, relative to the expected mean of the uphill eigenvalues, decreases as $N$ increases.  Consequently, landscape cosmologies may have the ability to soften  the inflationary $\eta$ problem by virtue of their dimensionality, independently of their detailed construction.

Many properties of the potential depend on just two parameters, the dimensionality $N$ and one additional parameter, $\gamma$ (defined in Ref. \cite{BBKS} and Section~\ref{notation}), where $0<\gamma<1$. When $\gamma$ is close to unity the potential is strongly {\em layered} \cite{auffinger2013complexity}; that is, the type of an extremum is increasingly correlated with the value of the potential at which it is found.  Consequently, for large $N$ and $\gamma$ close to unity the number of 1-saddles at positive values of the potential is vanishingly small.  However, when $\gamma$ is small 1-saddles  remain relatively plentiful at larger energies, including those at which typical  maxima are found. For any given potential, $\gamma$ can be computed from its power spectrum (again defined in Ref. \cite{Low2020} and Section~\ref{notation}); a small value of $\gamma$ is associated with a near scale-free spectrum. 

While Ref.~\cite{Low2020} analyzed the distribution of minima, this paper examines the properties of saddles which support either slow roll or topological inflation. Moreover,  Ref~\cite{Low2020} is concerned with only the relative signs of minima, a quantity that is unchanged by rescalings. In contrast, inflationary dynamics  depends on the physical magnitudes of the (downhill) slopes and heights of saddles and we will need to examine their properties more carefully. To set these free parameters consistently with broad physical expectations for a landscape potential we specify that the typical height of the landscape and its correlation length are both roughly Planckian. We consider three simple power spectra: Gaussian; and ``red'' and ``blue'' power-laws. In the latter cases the amplitude of the Fourier spectrum of the landscape potentials is tilted towards large and short scales in field space, respectively. Even these simple models  require subtle handling: if we define the average value of the potential energy $V^2$ to be unity, we will see that the height of a typical peak scales as $\sqrt{N}$. However, the inflationary consequences of a Gaussian landscape will depend on whether the typical magnitude of the potential is Planckian, or whether the typical \emph{peaks} are at the Planck scale. Conversely, we  find that saddles supporting simple slow-roll inflation \cite{Linde1982,Albrecht1982,Linde1983,Kinney2009} are rare -- but not vanishingly so -- and that many saddles at Planck scale energies would support topological inflation \cite{Vilenkin1994}.

The detailed implications of these results and actual predictions for inflationary observables would depend on further assumptions about the measure problem in the multiverse. However, we will be able to draw inferences about the  viability of different multiverses proposals, either via an appeal to self-consistency or because viable scenarios are so  rare that our specific universe cannot be plausibly produced within a given landscape proposal. Consequently, this work shows that quantitative analyses of some multiverse scenarios are in fact possible.

The rest of the paper is structured as follows: we will start by describing the properties of Gaussian random function and our notation in Section~\ref{defs}. In Section~\ref{general} we delineate the properties of the landscape that depend only on $N$ and $\gamma$; in Section~\ref{sec:properties} we show how to tie the dimensionless quantities in a random function to the physical scales of the landscape for three simple power spectra and compute the resulting properties of their 1-saddles. Section~\ref{sec:inflation} summarises the inflationary consequences of these findings, and we conclude in Section~\ref{sec:discuss}.

\section{Cosmology in a Gaussian Random Landscape}
\label{defs}

We consider a ``landscape'' scenario in which $N \sim {\cal{O}}(100)$ scalar fields $\phi_i$ (``moduli") interact via a potential $V(\phi)$, which is taken to be a statistically homogeneous and isotropic (in field space) Gaussian random function, with mean of zero. This Gaussian random function is unbounded and statistically invariant under the mapping $V\rightarrow -V$. This means for example that the probability  an extremum with a given value of $V$ is a local minimum is equal to the probability that an extremum at $-V$ is a local maximum,  or $P(\mathrm{min}|V) = P(\mathrm{max}|-V)$. The overall shape of a given realisation of a Gaussian random function is invariant under rescalings of the function and its arguments.  However, a Gaussian random function has a characteristic magnitude such that excursions well above this value are exponentially unlikely. This value defines a physical scale: roughly speaking it will be the ``height'' of the landscape. 

Our  analysis of the properties of random landscapes  builds on Ref.~\cite{Low2020}. The specific methodology is partially based upon an  $N$ dimensional generalization of the well-known  Bardeen, Bond, Kaiser and Szalay \cite{BBKS} treatment of the peaks of three dimensional random functions.

\subsection{Notation} \label{notation}

We begin by summarising our notation and laying out key definitions.

\begin{itemize}
    \item $n$-saddle: a saddle with $n$ downhill directions.
    \item $\eta_i$: the first derivative of $V$ in the $i$th direction, or $\partial V/ \partial \phi_i$. By definition, it is zero at an extremum.\footnote{This is consistent with Refs~\cite{BBKS,Low2020} but overlaps with the usual notation for the second slow roll parameter. However, since we focus on saddles with a single downhill direction the slow-roll  $\eta$ never has a subscript.} 
    
    \item $\zeta_{ij}$:  the Hessian of $V$, or the second derivative of $V$ in the $i,j$ directions,  $\partial^2 V/ \partial \phi_i \partial \phi_j$. 
    \item $N$: the dimension of the landscape; typically $N \sim O(100)$.
    \item $\xi(|\phi_1-\phi_2|)=\langle V(\phi_1)V(\phi_2)\rangle$: the two-point correlation function, a measure of how much knowledge of $V$ at $\phi_1$ reveals about its value at $\phi_2$. A low correlation implies little knowledge. For statistically homogeneous and isotropic Gaussian random landscapes only the distance between the two points matters; hence the absolute value sign. $\xi(|\phi_1-\phi_2|)$ is necessarily maximum at $\phi_1 = \phi_2$.
    \item $P(k)$: the power spectrum of the Gaussian random function. It can be defined as the Fourier transform of the correlation function, or:
    \begin{equation} \label{powspecdef}
        \xi(|\phi_1-\phi_2|)=\frac{1}{(2\pi)^N}\int d^Nk P(k) e^{ik\cdot(\phi_1-\phi_2)}    
    \end{equation}
        \item $\sigma_n$: moments of the power spectrum, defined by
    \begin{equation} \label{sigmadef}
    \sigma_n^2 = \frac{1}{(2\pi)^N}\int d^Nk (k^{2})^n P(k) \, .
   \end{equation}
   The first three moments satisfy  \cite{Low2020}:
    \begin{equation} \label{Correlations}
    \begin{aligned}
    \langle VV \rangle &= \sigma_0^2 \, ,\\
    \langle\eta_i\eta_j\rangle &=     \frac{1}{N}\delta_{ij}\sigma_1^2 \, , \\
    \langle V\zeta_{ij}\rangle &= -\frac{1}{N}\delta_{ij}\sigma_1^2 \, , \\
    \langle\zeta_{ij}\zeta_{lm}\rangle &= \frac{1}{N(N+2)}\sigma_2^2(\delta_{ij}\delta_{lm}+\delta_{il}\delta_{jm}+\delta_{im}\delta_{jl}) \, .
    \end{aligned}
    \end{equation}
In the above $\langle \rangle$ indicates an ensemble average and the two terms in each average are understood to be evaluated at the same point in field space. Consequently, $\sigma_0^2$ is  the variance of $V$, given that it has zero mean.
  \item $\nu \equiv V/\sigma_0$: the dimensionless potential.
 
    \item $\lambda_i$:  dimensionless eigenvalue(s) of the Hessian at an extremum in a basis in which $\zeta$ is diagonalized. $\lambda_i \equiv +\zeta_{ii}/\sigma_2$.\footnote{The positive sign likewise continues the convention of previous work \cite{Low2020}.}  Without loss of generality, we can order the eigenvalues such that $\lambda_1 \geq \lambda_2 \geq \ldots \geq \lambda_N$. At minima, all $\lambda_i \geq 0$; a 1-saddle  has $\lambda_1 \geq \lambda_2 \geq \ldots  \lambda_{N-1}  >0 >  \lambda_N$.
    \item $x_i$: a linear combination of the $\lambda_i$ \cite{Low2020}, defined as
    \begin{equation}
    \label{BasisTransform}
    x_1 = -\frac{1}{\sigma_2}\sum_i\zeta_{ii} = -\sum_i \lambda_i \, , \nonumber \quad
    x_n = -\frac{1}{\sigma_2}\sum_{i=1}^{n-1}\left(\zeta_{ii}-\zeta_{nn}\right) = -\sum_{i=1}^{n-1} \left(\lambda_i - \lambda_n \right),\,\, (2\leq n \leq N) \, .
    %\sigma_2x_3 &= -(\xi_{11}+\xi_{22}-2\xi_{33})\\
    %\sigma_2x_4 &= -(\xi_{11}+\xi_{22}+\xi_{33}-3\xi_{44})\\
    %\ldots
    \end{equation} 
    which will provide more compact expressions for the probability densities (Eq. \ref{Q}).
    \item $\gamma \equiv \sigma_1^2/(\sigma_0\sigma_2)$: In combination with $N$ this parameter fully specifies many of the properties of a Gaussian random potential \cite{Low2020,BBKS}. A potential whose power spectrum extends over a narrower range of scales will have a larger value of $\gamma$, as well as fewer minima at positive values of $V$.
\end{itemize}

\subsection{Probability Distributions of Extrema} \label{ReallyTheory}

The expected number density of {\em extrema} of a Gaussian random function is  \cite{BBKS,Low2020}
\begin{equation} \label{DensityOfPeaks}
\langle n_{{\rm extrema}} \rangle = A \int_{\lambda_1 \geq \lambda_2 \ldots \geq \lambda_N} G \times e^{-Q} \, d\nu \,d^Nx \, 
\end{equation}
where $A$ is a constant which absorbs the change of variables between $x_i$ and $\lambda_i$, plus the overall normalisation. The integrand in Eq. \ref{DensityOfPeaks} can be interpreted as an unnormalized likelihood for extrema, as a function of the $\lambda_i$ and $\nu$. The factor $G$ is\footnote{We have inserted absolute value signs for $\lambda_i$, which did not matter in Ref. \cite{Low2020} because all eigenvalues were positive then.}
\begin{equation}
G = \left(\prod_{i}^{N} |\lambda_i| \right)\left(\prod_{i<j} |\lambda_i-\lambda_j|\right),
\end{equation} 
where the second product drives the ``eigenvalue repulsion''  that is a common feature of random matrices. Finally, $Q$ has the generic form 
\begin{equation} \label{Q}
2Q = x_1^2 + \frac{(\nu-\gamma x_1)^2}{1-\gamma^2}+\sum_{n=2}^N\frac{N(N+2)}{2n(n-1)}x_n^2 +\frac{ N}{\sigma_1^2}  \sum_{i=1}^N  {\eta_i}^2+ \sum_{i,j;i > j}^N\frac{N(N+2)(\zeta_{ij})^2}{\sigma_2^2}
\end{equation}
At any extremum $\eta_i = 0$, so the fourth term in Eq. \ref{Q} vanishes.   We can pick axes such that $\zeta_{ij} = 0$ for $i \neq j$, eliminating the final term in $Q$. 

Eq. \ref{DensityOfPeaks} provides the number density of extrema. We restrict this to minima by modifying the region of integration to $\lambda_1 \geq \lambda_2 \ldots \geq \lambda_N > 0$. For 1-saddles, we adopt instead $\lambda_1 \geq \lambda_2 \ldots \geq 0 \geq\lambda_N$. These choices select different classes of extrema but $Q$ has only its first three terms in all cases. With this tweak the computational machinery developed in Ref.~\cite{Low2020} adapts smoothly to the current analysis. The properties of a landscape are defined by its power spectrum,  but key insight is provided by  $\gamma$, the only parameter that depends on the power spectrum in Eq. \ref{DensityOfPeaks}. It can be shown that $0 < \gamma < 1$ \cite{Low2020}; a landscape with smaller $\gamma$ has more ``fine structure'' than one with large $\gamma$, and a less sharply peaked distribution of minima as a function of $\nu$.

\begin{figure}[tb]
    \center{
        \includegraphics[width=.75 \textwidth]{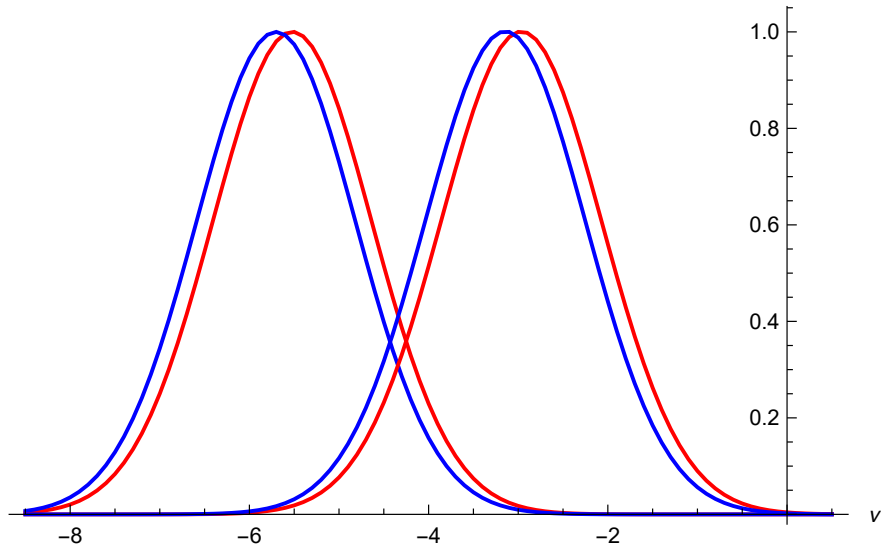}}\\
    
    \caption{%
       The likelihood for the most probable 1-saddle obtained by maximising $\cal{L}$  (Eq.~\ref{Likelihood})  with respect to the $\lambda_i$ separately for each value of $\nu$  (red) and the distribution found after integrating over the $\lambda_i$ (blue) using the Gaussian approximation. The left curve has $N=50$ and the right curve has $N=20$, with $\gamma = 0.5$, and arbitrary normalisation.  }
    \label{PeakV}
\end{figure}

The integral above implicitly defines a likelihood, 
\begin{equation} \label{Likelihood}
 {\cal L}(\lambda_1,\cdots\lambda_N;\nu,\gamma) =       \left(\prod_{i}^{N} |\lambda_i| \right)\left(\prod_{i<j} |\lambda_i-\lambda_j|\right) e^{-Q} \, .
\end{equation}
We can then compute quantities like the probability density of 1-saddles as a function of $\nu$,
\begin{equation} \label{normed}
    p(\nu|\gamma,\textrm{1-saddle}) d\nu = 
    \frac{\int {\cal L}(\lambda_1,\cdots\lambda_N;\nu,\gamma) d\lambda_1 \cdots  d\lambda_N}{\int^{-\infty}_\infty d\nu \int{\cal L}(\lambda_1,\cdots\lambda_N;\nu,\gamma) d\lambda_1 \cdots  d\lambda_{N}} d\nu
\end{equation}
where the integration volume is consistent with the constraints on the $\lambda_i$. In Ref.~\cite{Low2020} we showed how to construct a numerical, Gaussian approximation\footnote{We implemented this using the {\sc Mathematica} {\sc FindMaximum} command. A further transformation to map the $\lambda_i$ into the full real line allowed the use of the {\sc QuasiNewton} method (which does not handle maximisation on a constrained domain, as currently implemented),  which improved performance dramatically for larger $N$. This is effectively the Broydon-Fletcher-Goldfarb-Shanno  algorithm, which is also implemented in {\sc SciPy} \cite{2020SciPy-NMeth}.} that marginalises over uninteresting parameters to obtain quantities such as the distribution of peaks as a function of $\nu$. However, up to their normalisation, these distributions are well-approximated by the maximal value of the ``raw'' likelihood for given value(s) of the parameter(s) of interest, as illustrated in Fig.~\ref{PeakV}, and these are simpler and numerically cheaper to obtain.  Given that this analysis deals primarily with qualitative results, we work with the maximised likelihoods in what follows, rather than the similar distributions that follow from integrating out extraneous variables. 

\section{General Landscape Properties}
\label{general} 

\begin{figure}[tb]
    \centering
    \includegraphics[width=.9 \textwidth]{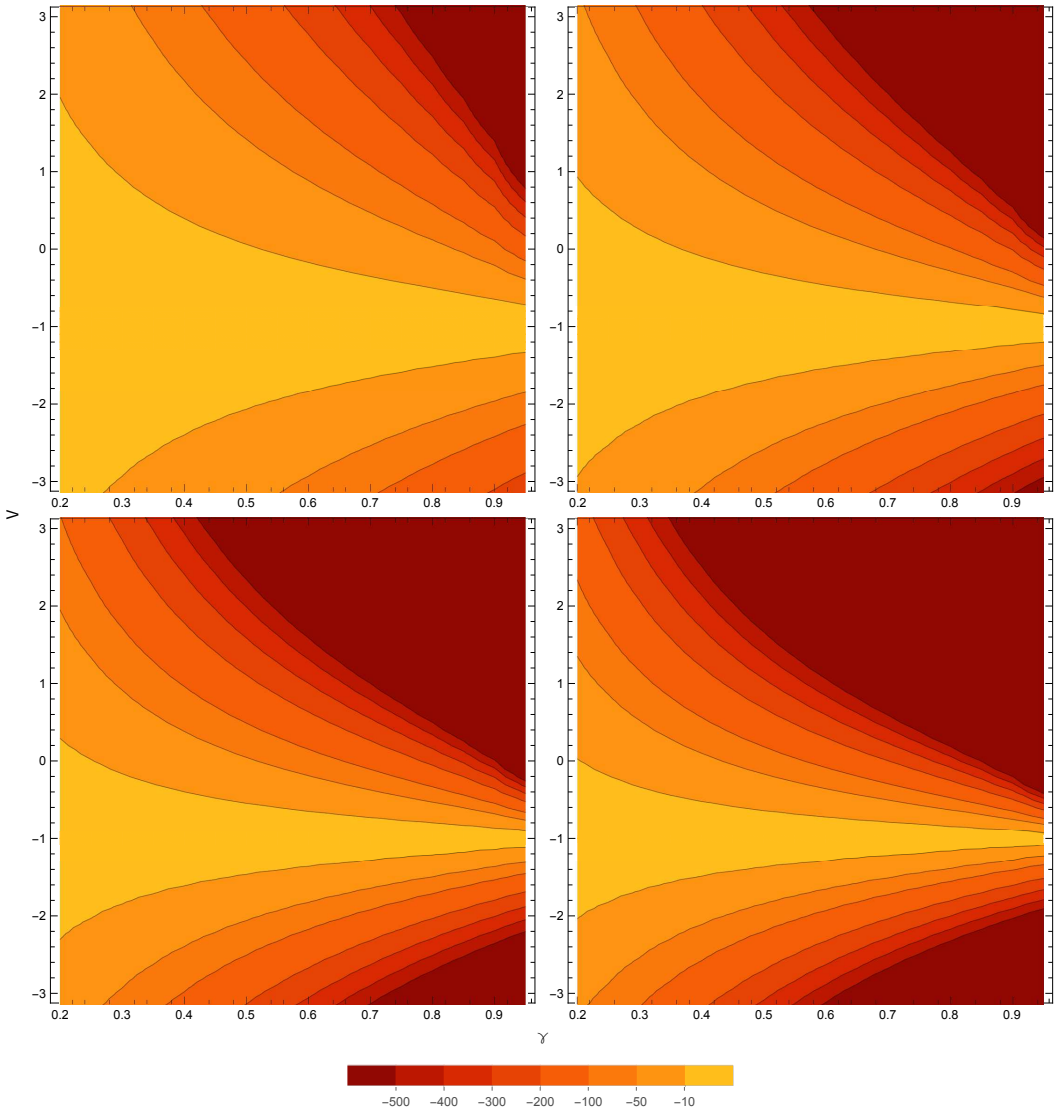}
    \caption{%
        Likelihood of a 1-saddle as a function of $\gamma$ and $V/|V_{\mathrm{max}}|$, where $V_{\mathrm{max}}$ is the value of $V$ at which the most 1-saddles are located. The plots are for $N$ of 50 (top left), 100, 200 and 300 (bottom right). The contours show $\log_{10}\mathcal{L}$; the lowest lying contour is at $10^{-500}$. As $N$ and $\gamma$ increase the fraction of 1-saddles with $V > 0$ decreases.
    }
    \label{TopologicalFig}
\end{figure}

We proceed by determining the region of the parameter space ($\gamma$, $N$ and the total number of vacua) in which we can expect any 1-saddles to be ``above the waterline". If $10^{500}$ is a rough rule of thumb for the number of minima we  expect a landscape to support, the expected number of 1-saddles is $N \times 10^{500}$ \cite{Easther2016}, or about $10^{502}$ if $N$ is of order $100$, which is close enough to $10^{500}$ to be functionally equivalent.\footnote{We are primarily interested in large differences in logarithmic probabilities; this also means that residual errors in our maximisation algorithms \cite{Low2020} can be safely ignored.}  The 1-saddles have a single downhill direction, so they are ``almost minima'', and will be rare for parameter choices at which minima are rare. Fig.~\ref{TopologicalFig} shows the relative likelihood of 1-saddles as a function of $V$ and $\gamma$, for four different values of $N$; we see that for $N\gtrsim 200$ there is a large range of values of $\gamma$ for which we would not expect any 1-saddles at positive values of $V$ in simple landscape scenarios, mirroring the result previously found for minima  \cite{Low2020}.

Looking at the eigenvalue distributions, $\nu$ appears just once in Eq. \ref{Likelihood} -- in the factor $Q$ (Eq. \ref{Q}). Therefore the maximum likelihood value of $\nu$ for any 1-saddle is
\begin{equation}
\nu = \gamma x_1 \, ,
\end{equation}
irrespective of the value of $x_1$. Maximising the likelihood with respect to $x_i$ we obtain both the most likely value of $\nu$ at which to find a 1-saddle as well as the most likely eigenvalues. At the peak likelihood, the eigenvalue distribution is independent of $\gamma$ \cite{Low2020}. By definition, $x_1$ is the sum of the eigenvalues, so $x_1 = N \bar{\lambda}$ where $\bar{\lambda}$ is the mean of the dimensionless eigenvalues of the Hessian matrix. These are related to the mass terms of the scalar fields in the vicinity of the extremum, or $m_i^2 = \sigma_2 \lambda_i$. This allows us to write down a relationship between the value of the potential and the average mass of the fields at the most probable extrema
\begin{equation} \label{gammaN}
  %     \frac{\nu}{N\bar{\lambda}} =  \frac{\sigma_2}{\sigma_0}\frac{ V}{   N\bar{m}^2} = \gamma
\bar{m}^2 =  \frac{\sigma_2}{\sigma_0}\frac{ V}{   N\gamma}
\end{equation}
where $\bar m^2 = \sigma_2 \bar \lambda$.

\begin{figure}[tb]
    \center{
    \includegraphics[width=.48 \textwidth]{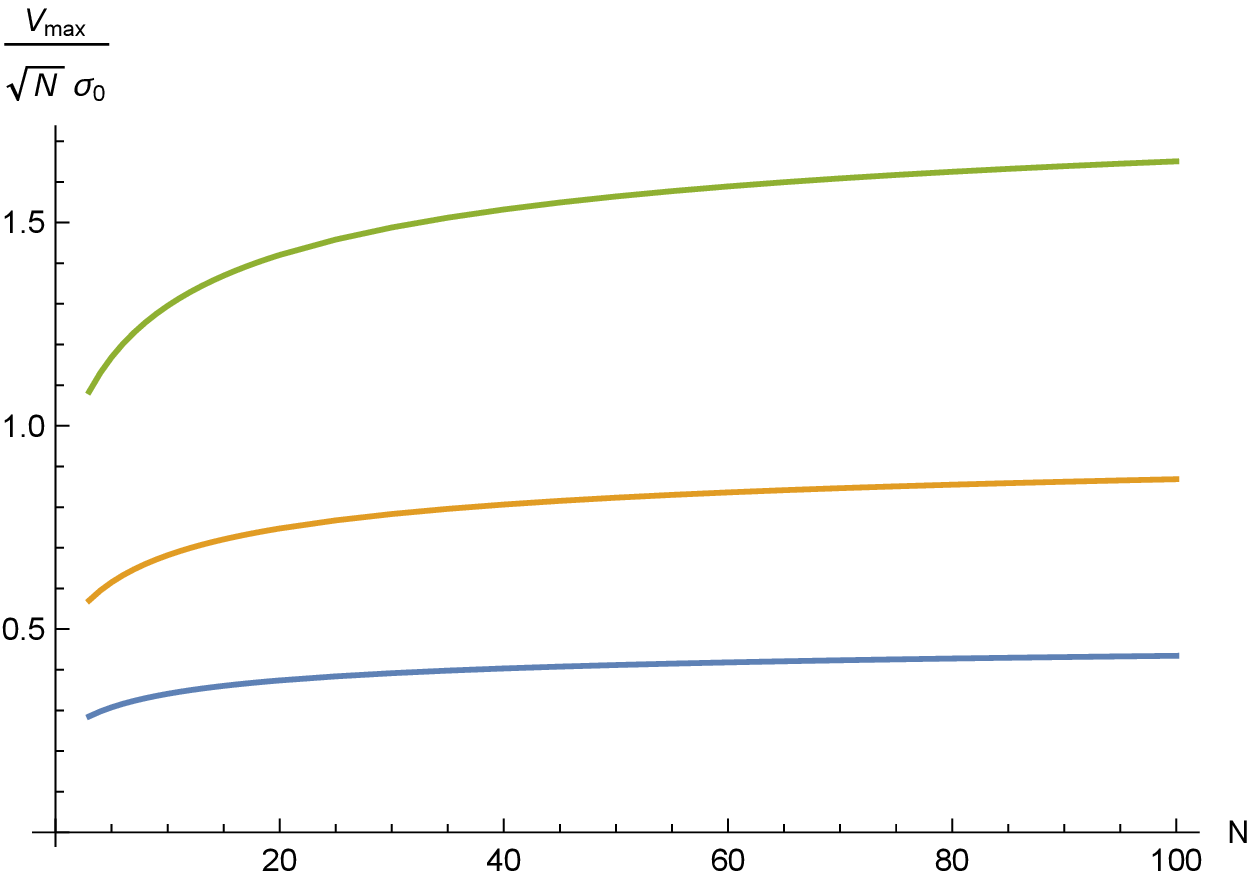} \hfill
      \includegraphics[width=.48 \textwidth]{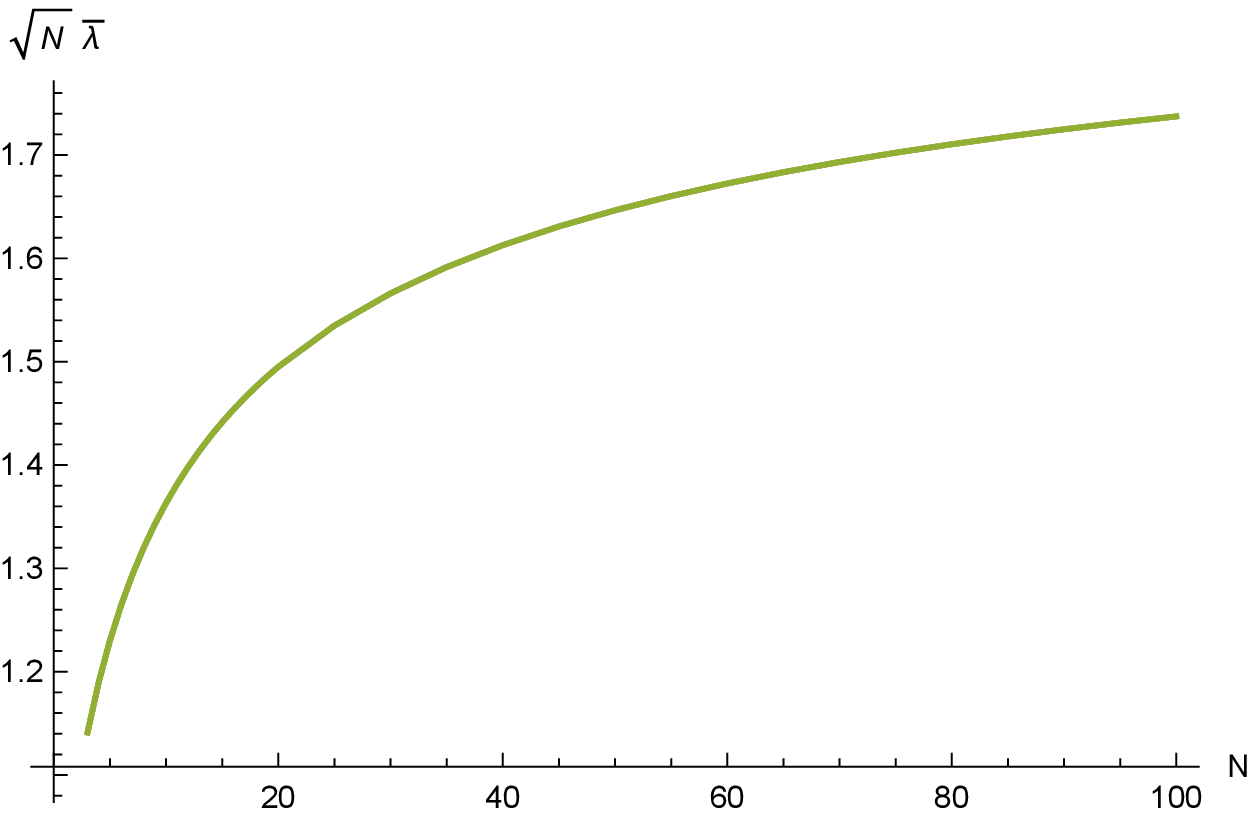}} %Generated by "Generating scaling relations.nb"
        \caption{%
        (Left) The value of $V$ at the point of maximum likelihood for maxima is shown as a function of $N$, for (from bottom to top)  $\gamma=0.25$, $0.5$ and $0.95$. The results have been divided by $\sqrt{N}$ to illustrate their scaling behaviour.
        (Right) The mean dimensionless eigenvalue $\bar{\lambda}$ at the  most probable 1-saddle is plotted as a function of $N$; it is independent of $\gamma$ and scales as approximately $1/\sqrt{N}$.
    }
    \label{Vscaling}
\end{figure}

A landscape has two key scales: a height; and a ``width'', which can be defined in terms of a correlation length. The average height of the landscape is related to  $\langle V^2 \rangle = \sigma_0^2$ (Eq. \ref{Correlations}). The string landscape is expected to reach Planckian energies, which suggests that $\sigma_0 \approx M_P^4$. The number density of \emph{maxima} is a function of $\nu$. In Fig. \ref{Vscaling} (left) we show the most probable value of $\nu$ at a maximum. We see that this value scales as $\sqrt{N}$ and the constant of proportionality increases with $\gamma$. Consequently, setting $\sigma_0$ to $M_P^4$  implies that most maxima of the potential  will be classically inaccessible (i.e. at $V > M_{P}^4$) or, more allegorically, the mountaintops of the landscape will always be in the clouds. Likewise, given the  $V\rightarrow -V$ symmetry, this scenario would have few classically stable minima. Alternatively, choosing $\sigma_0 \sim M_P^4/\sqrt{N}$ ensures that a typical maximum has $V\sim M_{P}^4$ so most maxima are classically accessible.\footnote{Note that for this choice the root-mean-square energy density of the landscape is $M_P^4/\sqrt{N}$ per Eq. \ref{Correlations}.} Looking at Fig. \ref{Vscaling} (right) we see that the mean dimensionless eigenvalue at 1-saddle scales (roughly) as $1/\sqrt{N}$.  

\begin{figure}[tb]
    \center{
    \includegraphics[width=.98  \textwidth]{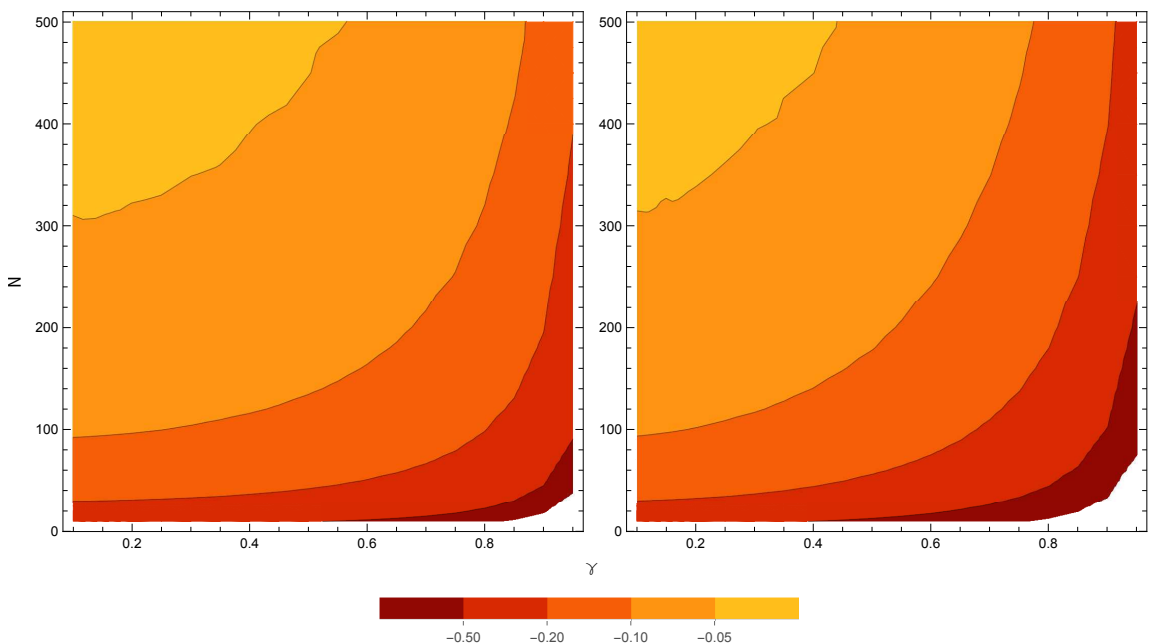} }
    
    \caption{%
       Ratios of the downhill eigenvalue to the average of the uphill eigenvalues, as a function of $N$ and $\gamma$, for typical 1-saddles. The lefthand plot shows the ratio at $V=0$; the righthand plot at $V=|V_{\mathrm{max}}|$, where $V_{\mathrm{max}}$  is the most likely value of $V$ for a 1-saddle. For small $\gamma$ and large $N$ these ratios can be on the order of $10^{-2}$.  }
    \label{eigenratio}
\end{figure}

\begin{figure}[tb]
    \center{
    \includegraphics[width=.45 \textwidth]{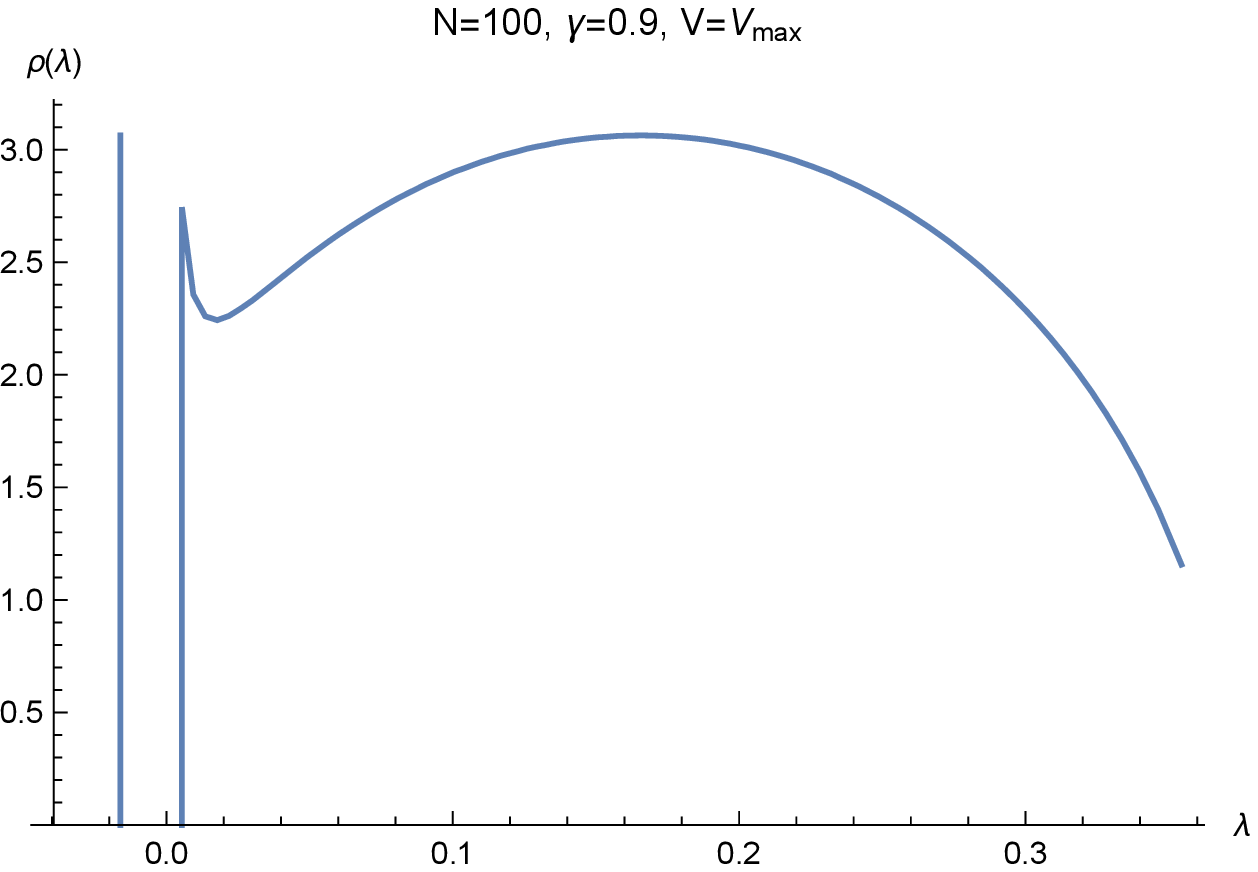} \hfill
    \includegraphics[width=.45 \textwidth]{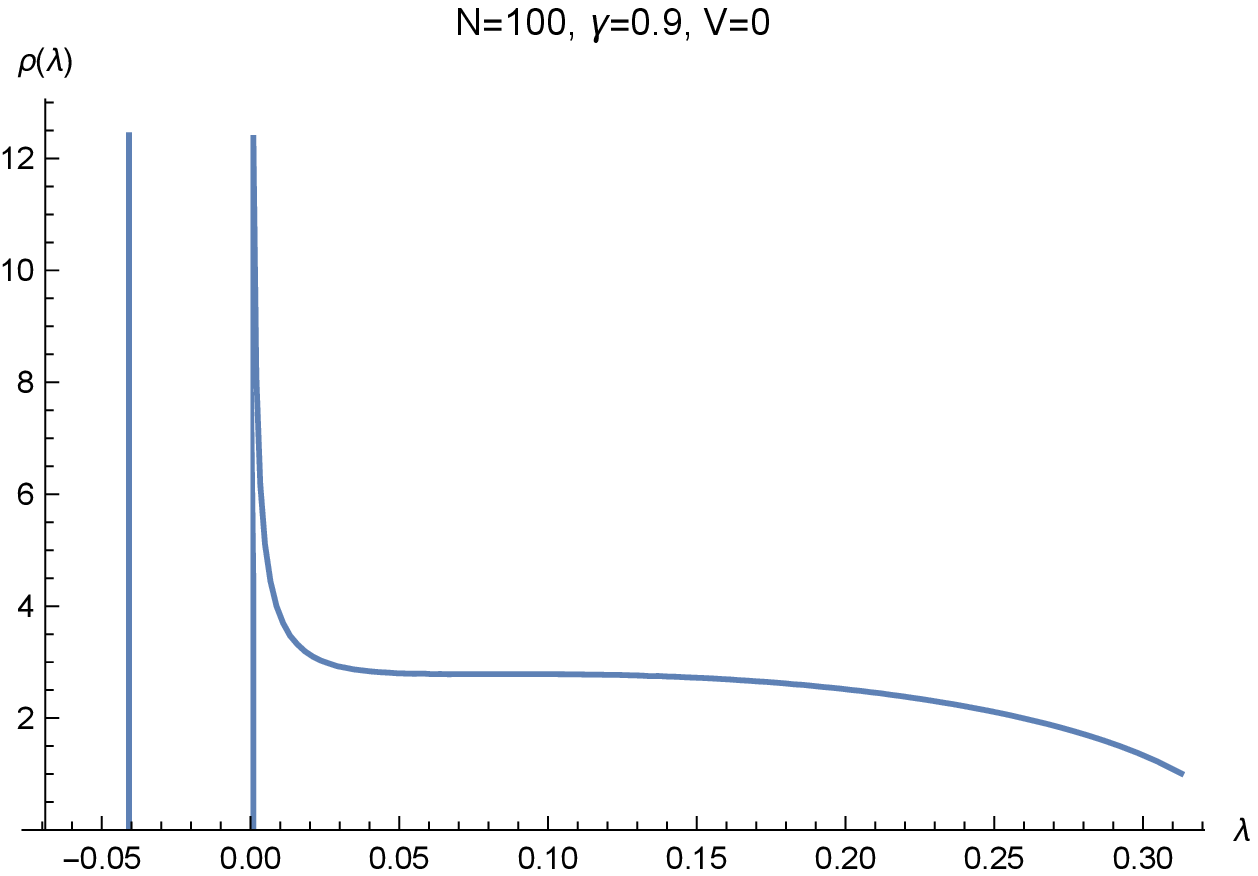} \\ 
\vskip 5mm
 \includegraphics[width=.45 \textwidth]{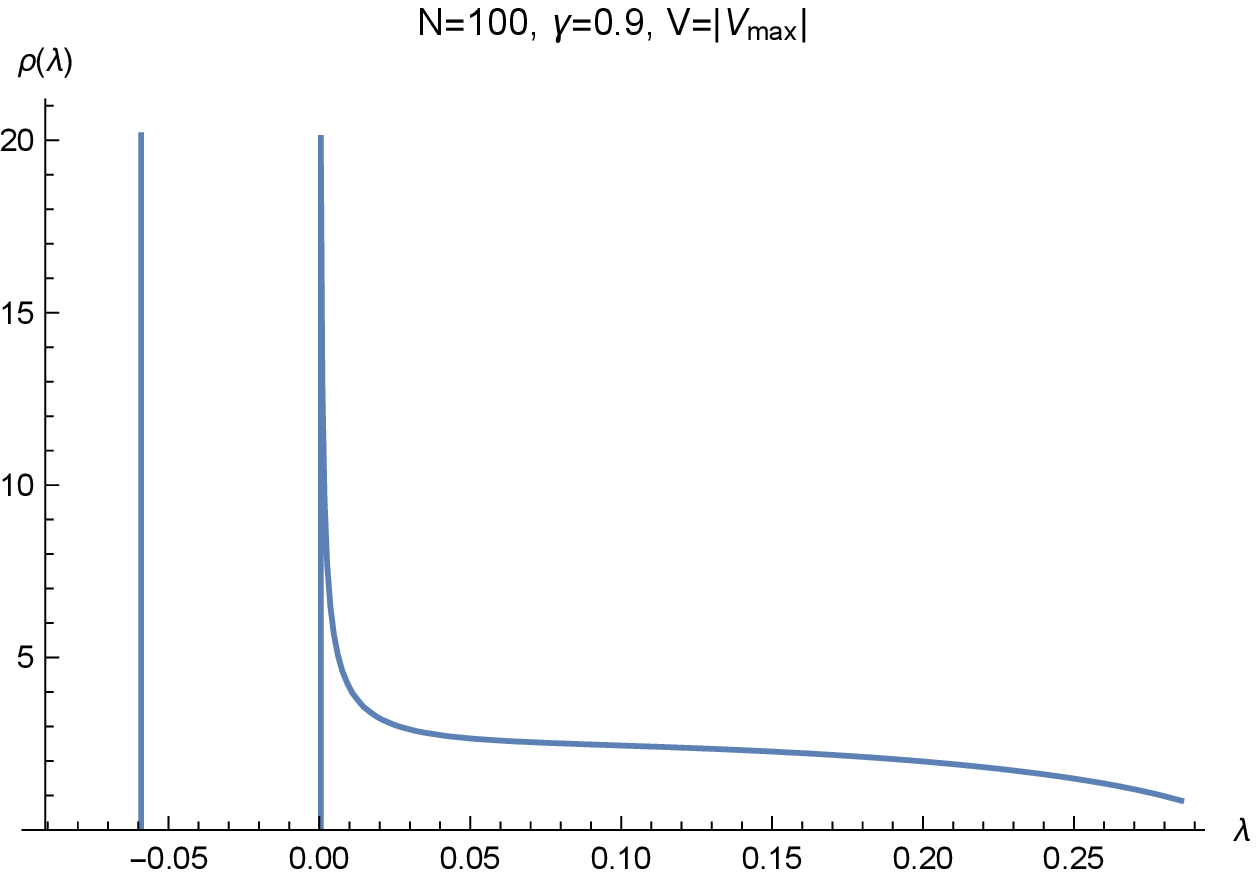} \hfill
    \includegraphics[width=.45 \textwidth]{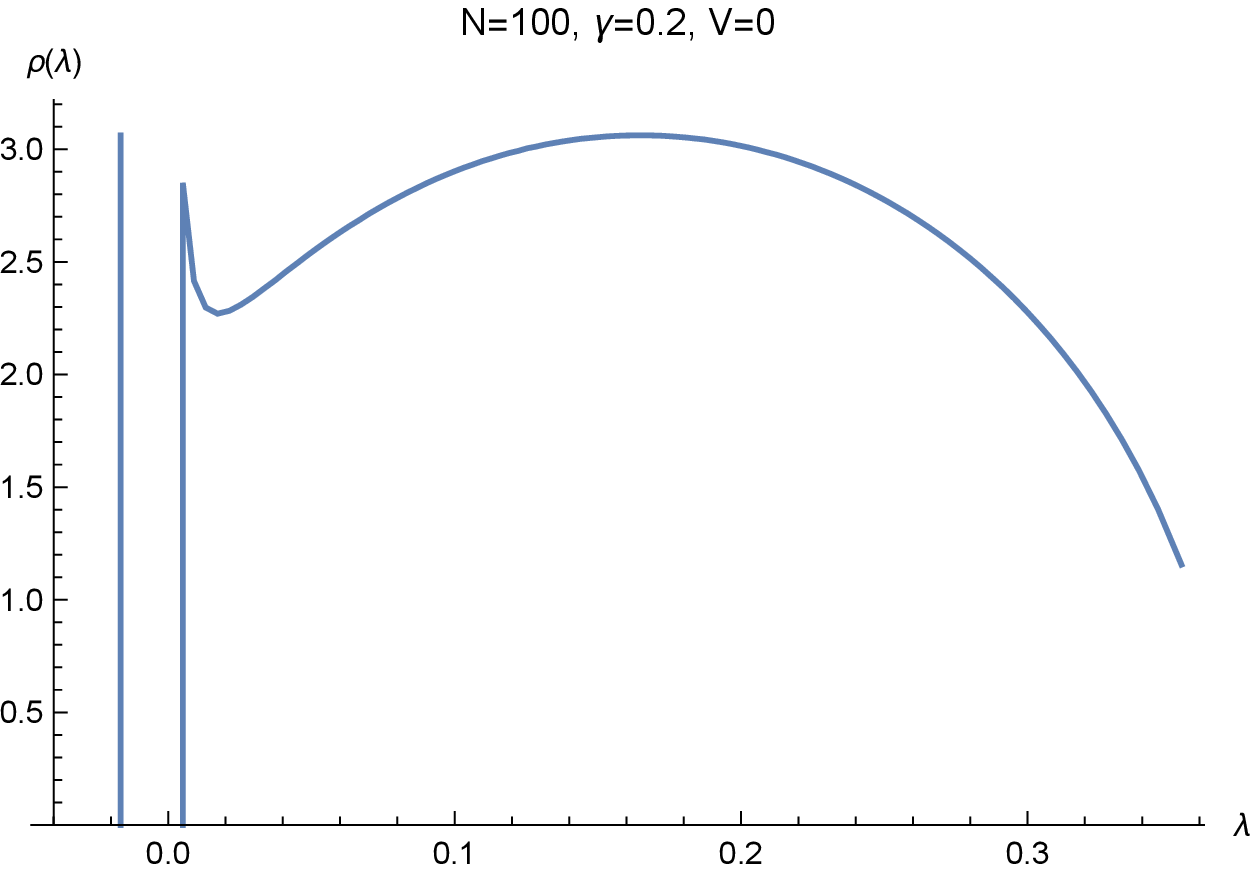}}
    
    \caption{%
   Approximate most probable eigenvalue distributions for 1-saddles. The upper left, upper right and lower left plots are for $V_{\mathrm{max}}$, the most likely value of $V$ for a 1-saddle, $V=0$ and $V=|V_{\mathrm{max}}|$. Because 1-saddles are `almost minima', $V_{\mathrm{max}}$ is negative. It can be seen that as $V$ increases the positive eigenvalues accumulate near zero, while the downhill eigenvalue increases (i.e., the downhill direction becomes progressively steeper). Conversely, on the lower right we show $\rho(\lambda)$. In this case the distribution is largely independent of $V$; the distributions at $V=\pm|V_{\mathrm{max}}|$ (which are not plotted) are essentially identical.}
    \label{eigendens}
\end{figure}

Figure \ref{eigenratio} highlights an interesting feature of the 1-saddles: the ``downhill'' eigenvalue (relative to the average uphill eigenvalue) at the most likely 1-saddle decreases as $N$ is increased or $\gamma$ decreased. This is the flip-side of eigenvalue repulsion -- too large a gap between the smallest positive eigenvalue and the negative eigenvalue is disfavoured by the third term in Eq. \eqref{Q}.  In Fig.~\ref{eigendens} we show an approximation to the most probable distribution of all the eigenvalues\footnote{These plots  are constructed from the differences between the most likely eigenvalues; at $\lambda_i$ the probability density is roughly proportional to $1/(\lambda_{i+1} -\lambda_i)$ and we show the negative eigenvalue as an effective $\delta$-function. Consequently, as plotted, $\rho(\lambda)$ can stop short of $\lambda=0$ and does not intersect the $x$-axis at large $\lambda$.} for 1-saddles at representative choices of $N$ and $\gamma$. For parameter combinations where 1-saddles are very unlikely the positive eigenvalues ``pile up'' at zero, and the downhill eigenvalue becomes larger, while for more probable 1-saddles we get a Wigner semicircle with deviations at very small values of $\lambda$ (compare Fig. 8 in \cite{Low2020} and Figs. 8 and 9 in \cite{Yamada2018}).    Note that this is solely a statement about the {\em relative} sizes of the most likely eigenvalues so it is thus independent of rescalings of the potential and fields.   

In the absence of other information, one might expect that the single downhill direction of a 1-saddle would have a slope similar  to a typical uphill direction. The fact that this is not the case is  a non-trivial  property of  Gaussian random landscapes. Physically, this  is a   mechanism by which random landscape cosmologies could address the  $\eta$-problem, as this amounts to the need for a ``small mass'' relative to a fundamental scale which, on the face of it, represents a tuning~\cite{Copeland:1994vg,Baumann2015}.

\begin{figure}[tb]
    \center{
    \includegraphics[width=.55 \textwidth]{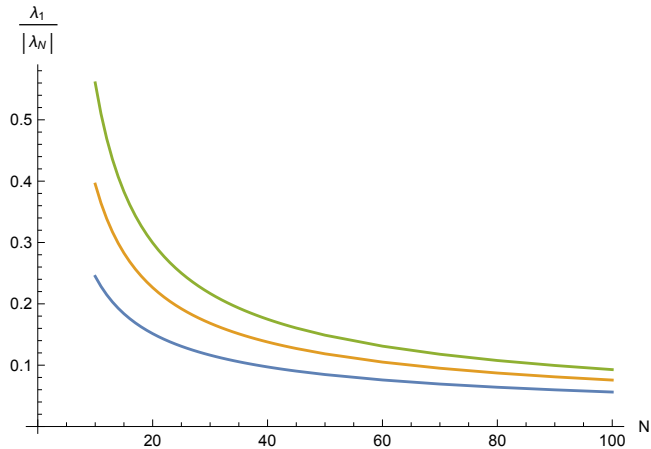}}
    \caption{%
      We  plot the ratio of the largest positive and largest (or only) negative eigenvalues for saddles with 1,2 or 3 downhill directions (bottom to top), $\nu=0$ and $\gamma=0.5$. The ratios scale roughly as $1/\sqrt{N}$. Note that $\nu=0$ is a good approximation to potential values needed for a ``realistic'' inflationary epoch, given the constraints on the gravitational wave background. 
    }
    \label{nsaddles}
\end{figure}

In this paper we have chosen to focus on 1-saddles. This is partly a pragmatic decision since  effective single-field models are much simpler to analyse. However, the inflationary properties of an $n$-saddle  depend strongly on its largest downhill eigenvalue and Fig.~\ref{nsaddles} shows the ratios of the largest positive eigenvalue to the largest negative eigenvalue for 1-, 2- and 3-saddles, a quantity which increases with the number of downhill directions. For  $V>0$  $n$-saddles will be more numerous than 1-saddles. However, for most cases of interest successful inflation will require a rare saddle at which  the steepest  eigenvalue is atypically small -- and getting several eigenvalues that are all smaller than  this critical threshold will require a more extreme  excursion than with a single eigenvalue, providing further motivation for our choice.

\section{Saddles, Scales and Dimensions} 
\label{sec:properties}
We now consider inflation in this prototype landscape. Both slow-roll and topological inflation require saddles that satisfy flatness criteria, which in a single field approximation are described in terms of the slow roll parameters
\begin{align}
    \epsilon &= \frac{M_{P}^2}{2}\left(\frac{V'}{V}\right)^2 \, , \\
    \eta &=M_{P}^2 \frac{V''}{V} \,  . \label{EtaDef}
\end{align} 
where $M_P$ is the Planck Mass, and $V$ is the value of the potential. For inflation to occur, both parameters must be small.\footnote{Strictly speaking, only $\epsilon$ needs to be small for inflation, but sustained inflation with $\eta > 1$ requires a special potential.} At a saddle, $\epsilon$ is guaranteed to vanish because $V'=0$.  In our notation  $\eta$ is  
\begin{equation} \label{eq:etaval}
    \eta = M_{P}^2 \frac{\lambda \sigma_2}{\nu \sigma_0}
\end{equation}
In other words, $\eta$ depends on the moments of the power spectrum $\sigma_0$ and $\sigma_2$, the dimensionless downhill eigenvalue $\lambda$, and the dimensionless potential value $\nu$. 

To assess the likely value of $\eta$ we need to first specify a power spectrum. We are looking for small  $\eta$  and a bounded \textit{correlation length}, $L$, which is the typical separation in field space at which the values of the landscape are effectively uncorrelated.  Loosely speaking, knowledge of the landscape at a given point will provide very little information about the landscape at points further away than $L$. On physical grounds, other than at special points, it is unlikely that the landscape is correlated on scales much larger than $M_P$ -- it would imply many sub-Planckian terms yield a consistent sum on super-Planckian scales. Consequently, our choices for the free parameters in the power spectra will need to reflect this expectation. 

Generally, $L$ is inversely correlated with $\eta$. From the definitions (see Section \ref{notation}), it is clear that $\sigma_0^2 = \xi(0)$, the peak of the correlation function. Differentiating Eq. \ref{powspecdef} with respect to $\phi$, we also have that $\sigma_1^2 \propto - d^2\xi/d\phi^2$ (evaluated at the point $|\phi_1-\phi_2|=0$). A correlation function that decreases quickly (corresponding to larger $d^2\xi/d\phi^2$ and thus larger $\sigma_1$) would also have a shorter correlation length. On the other hand, $\eta$ is proportional to $\sigma_2$. For fixed $\gamma = \sigma_1^2/\sigma_0\sigma_2 < 1$, a decrease in $\sigma_2$ requires a corresponding decrease in $\sigma_1$,\footnote{Note that $\sigma_0$ cannot change because it is fixed to the root mean square energy of the landscape -- $M_P^4$ or $M_P^4/\sqrt{N}$.}  and thus an increasing correlation length. 

Nonetheless, this is not a hard no-go theorem, and $\eta$ only needs to be of the order $10^{-2}$, rather than exponentially suppressed.  To get a sense of how different power spectra control expectations for $\eta$ we  examine three representative examples --  a Gaussian spectrum; and red and blue power-law spectra. We can compute $\sigma_0$ and $\sigma_2$ for these cases, from which we obtain $\eta$  as a function of $V$ and the other parameters. 

\subsection{Gaussian Power Spectrum}

The Gaussian power spectrum\footnote{Note that a Gaussian random function is one where the value of the function at any point is drawn from a Gaussian distribution. In this case the power spectrum is {\em also} Gaussian.} can be defined via a Gaussian correlation function  
\begin{equation} \label{CorrFunc}
\xi(\phi)=U_0^2 e^{-\phi^2/2L^2} = \frac{1}{(2\pi)^N} \int d^Nk P(k) e^{ik \cdot \phi}
\end{equation}
where $U_0$ is the amplitude of the Gaussian and $L$ is the correlation length. The Fourier transform of a Gaussian is a Gaussian, so the power spectrum is also Gaussian. The $\sigma_n^2$ are~\cite{Yamada2018}
\begin{equation}
\sigma_n^2 = \frac{2^n U_0^2}{L^{2n}} \frac{\Gamma[n+\frac{N}{2}]}{\Gamma[\frac{N}{2}]} \, ,
\end{equation}
which gives
\begin{eqnarray}
    \gamma &=& \sqrt{\frac{N}{N+2}}  \, ,\\
\frac{\sigma_2}{\sigma_0} &=& \frac{1}{L^2} \sqrt{N(2+N)} \, .
\label{eq:gaussianetaratio}
\end{eqnarray}
These two expressions depend only  the number of dimensions and the correlation length, $L$, which is a free parameter. However,  a landscape with a Gaussian power spectrum has a vanishingly small number of 1-saddles at positive values of $V$, since $\gamma$ is not a free parameter and tends to large values as $N$ increases. For $N = 100$, $\gamma \approx 0.990$. Given  the results of the previous Section (Fig. \ref{TopologicalFig}) we expect that the odds of a given saddle being  ``above the waterline" are small. For this specific example they about 1 in $10^{1197}$, so even with $10^{502}$ 1-saddles the odds are roughly 1 in $10^{695}$  a single 1-saddle will be found at positive $V$.  This result is fully analogous to the conclusions previously drawn in Ref. \cite{Low2020} about the viability of a landscape with a Gaussian power spectrum based on the number of minima. 

\begin{figure}[tb]
    \center{\includegraphics[width=0.55 \textwidth]{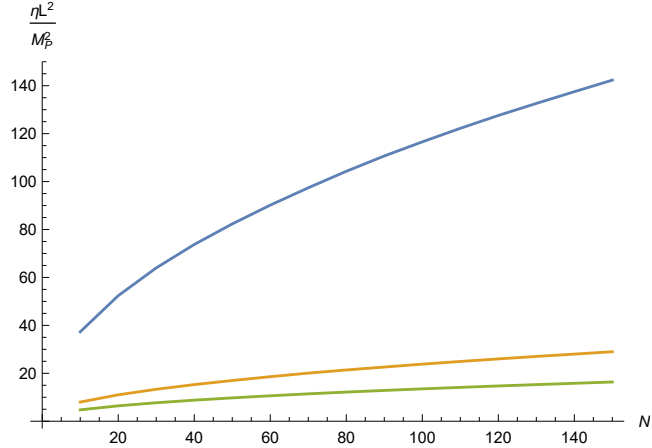}}
    \caption{%
        The most probable value of the slow-roll inflation parameter $\eta$ multipled  by $L^2/M_P^2$, as a function of the number of dimensions $N$ for a Gaussian power spectrum. We plot this quantity for three different $\nu$:  $\nu = 0.1, \nu = 0.5$, and $\nu = 0.9$ (top to bottom). 
    }
    \label{GaussianEta}
\end{figure}

As an illustrative exercise, we can still determine the most likely value of $\eta$ as a function of $V$.  Using Equations~\ref{eq:etaval} and  \ref{eq:gaussianetaratio} and the most probable downhill $\lambda$ calculated by maximizing Eq. \ref{Likelihood}, we plot the most probable values of $\eta$ as a function of $N$ in Fig. \ref{GaussianEta}. As expected, with larger $\nu$ the most probable value of $\eta$ is smaller. Of course, choosing a large value of $L$ would drive $\eta$ toward small values. However, this would be an unphysical choice for the correlation length that effectively smuggles a tuning into the setup. 

\subsection{Power-law power spectrum}
The power-law power spectrum is
\begin{equation}
    P(k) = A k^{-n}
\end{equation}
where $A$ is a constant. The integral over all $k$ is divergent so we impose cutoffs; for the ``red'' case with  $P(k) = 0$ for $k < k_{cut}$, while for the opposite ``blue'' case $P(k) = 0$ for $k > k_{cut}$. The power-law power spectrum  thus has three parameters: $A$, $n$, and $k_{cut}$; $A$ sets the overall scale, while $n$ and $k_{cut}$ control the properties of the landscape.  

\begin{figure}[tb]
  \centering
  \includegraphics[width=0.55 \linewidth]{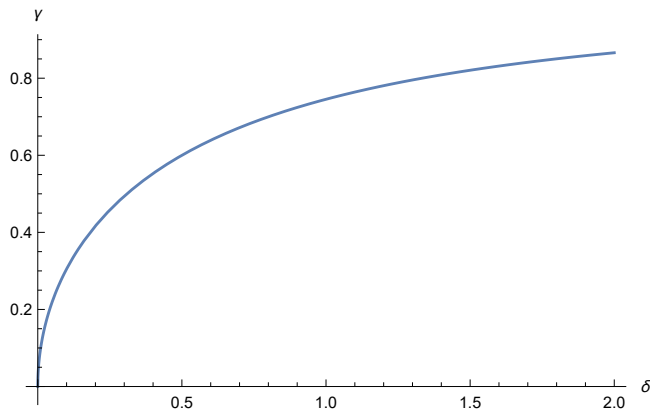}
  \caption{We plot $\gamma$ for the red power-law power spectrum $P(k) = Ak^{-n}$ as a function of $\delta$. We see that $\gamma$ can take any value, and for sufficiently small $\delta$ it will be close to zero.}
  \label{PowerLawPowerSpectrumGamma}
  \end{figure}

For a red cutoff, the moments of the power spectrum are:
\begin{align} \label{RedPowerLawSigmas}
    \sigma_0 &= \frac{1}{(2\pi)^{N/2}} \sqrt{\frac{A k_{cut}^{N-n}}{n-N}} \quad (n > N)  \, ,\\
    \sigma_1 &= \frac{1}{(2\pi)^{N/2}} \sqrt{\frac{A k_{cut}^{2+N-n}}{n-N-2}} \quad (n > N + 2) \, , \\
    \sigma_2 &= \frac{1}{(2\pi)^{N/2}} \sqrt{\frac{A k_{cut}^{4+N-n}}{n-N-4}} \quad (n > N + 4) \, ,
\end{align}
from which we find
\begin{equation} \label{redcutoffgamma}
\gamma = \frac{\sqrt{(n-4-N)(n-N)}}{n-N-2} \, .
\end{equation}
This equation is valid only if $n > N + 4$. We see that $\gamma$ does not depend on $k_{cut}$ but does depend on $n$. Writing $n = N + 4 + \delta$, $\gamma = \sqrt{\delta(4+\delta)}/(2+\delta)$ and the resulting relationship is shown in Fig. \ref{PowerLawPowerSpectrumGamma}. As can be seen, while $\gamma$ is fixed by $N$ for the Gaussian power spectrum, a power-law can generate all values of $\gamma$, but smaller values of $\gamma$ require a near-flat spectrum. 

\begin{figure}[tb]
    \center{
    \includegraphics[width=\textwidth]{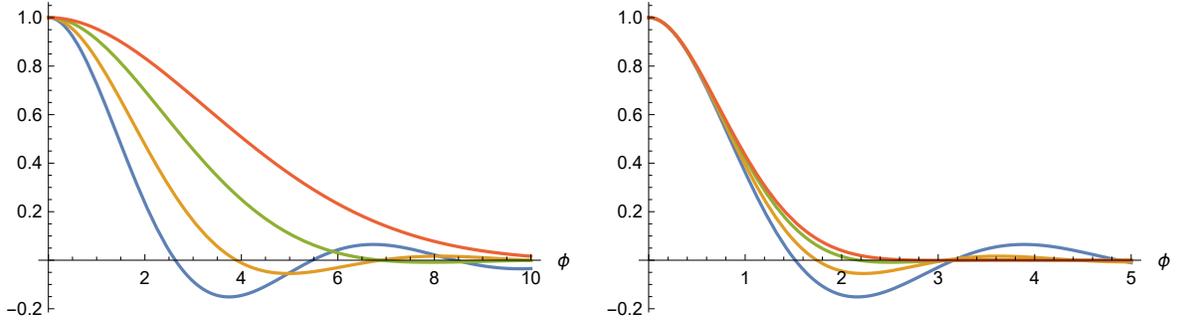} }
        \caption{%
       The normalized correlation function ($y$-axis) as a function of $\phi$, for (from bottom to top) $N = 3$, $5$, $10$ and, $20$ with $\delta = 0.1$.  Left: the normalized correlation function with $k_{cut} = 1$. Right: the normalized correlation function with $k_{cut} = \sqrt{N}$.
    }
    \label{kmin}
\end{figure}

Using the moments of the power spectrum, we can calculate $\eta$ 
\begin{equation} \label{RedPowerLawEtaDef}
    \eta = M_P^2 \frac{\lambda}{\nu} k_{cut}^2 \sqrt{\frac{4+\delta}{\delta}} \, .
\end{equation}
We set $k_{cut}$ by examining the correlation function. By definition, the correlation function is the Fourier transform of the power spectrum, Eq. \ref{powspecdef}. Performing the integral yields 
\begin{equation}
    \xi(|\phi_1|)=\frac{A}{(2\pi)^N}\int_{k_{cut}}^{\infty} dk k^{N-n-1}(\phi_1 k)^{1-N/2} J_{N/2-1}(\phi_1 k)     
\end{equation}
where $J$ is the Bessel function of the first kind, and we have used translation invariance to set $\phi_2 = 0$. This integral can  be evaluated (via Mathematica) to give
\begin{align}
     \xi(|\phi|) =& \frac{A}{(2\pi)^N}\frac{1}{\Gamma[n/2]}2^{-n-N/2}k_{cut}^{-n}\phi^{-N}\Gamma\left[\frac{N-n}{2}\right]\times \nonumber \\ & \left\{2^N(k_{cut}\phi)^n-2^n(k_{cut}\phi)^N \Gamma[n/2]   _1\tilde{F}_2\left(\frac{N-n}{2};\frac{N-n}{2} + 1,\frac{N}{2};-\frac{1}{4} k_{cut}^2 \phi ^2\right) \right\}
\end{align}
where $_1\tilde{F}_2$ is a regularized generalized hypergeometric function. Plots of this function for representative parameter values are given in Fig.~\ref{kmin}; the correlation length is  roughly the distance to the point where $\xi$ becomes sensibly zero, and this grows with $N$ if $k_{cut}$ is constant. On the other hand, with $k_{cut} = \sqrt{N}$ the correlation length tends to a fixed value near unity\footnote{For this specific choice is it is around 2.5, but the parametric scaling with $N$ is  sufficient for our needs.}  at large $N$, consistent with expectations for a landscape. We therefore adopt $k_{cut} = \sqrt{N}/M_{P}$ in what follows. 

\begin{figure}[tb]
    \center{
    \includegraphics[width=.49\textwidth]{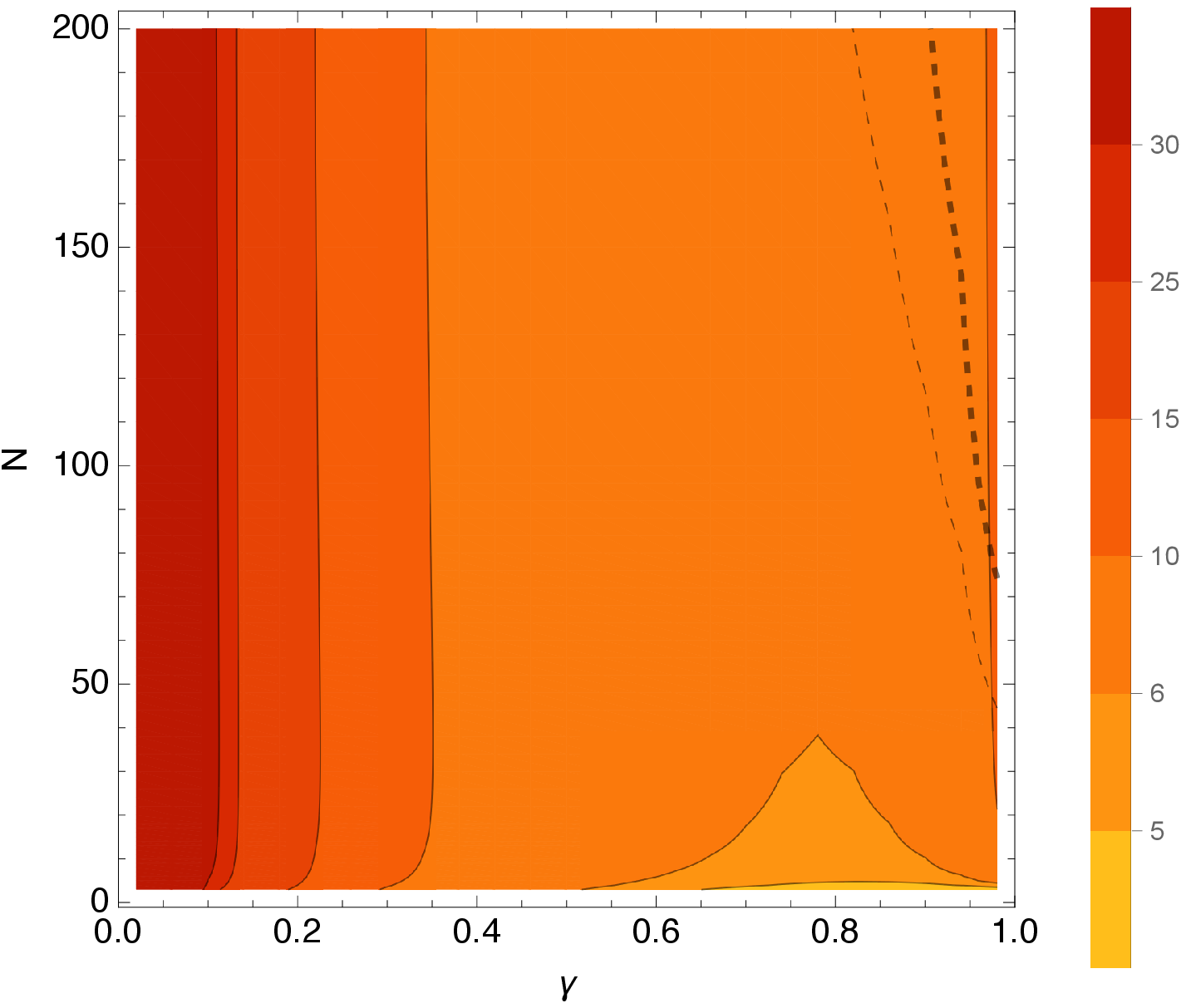} \hfill 
    \includegraphics[width=.49\textwidth]{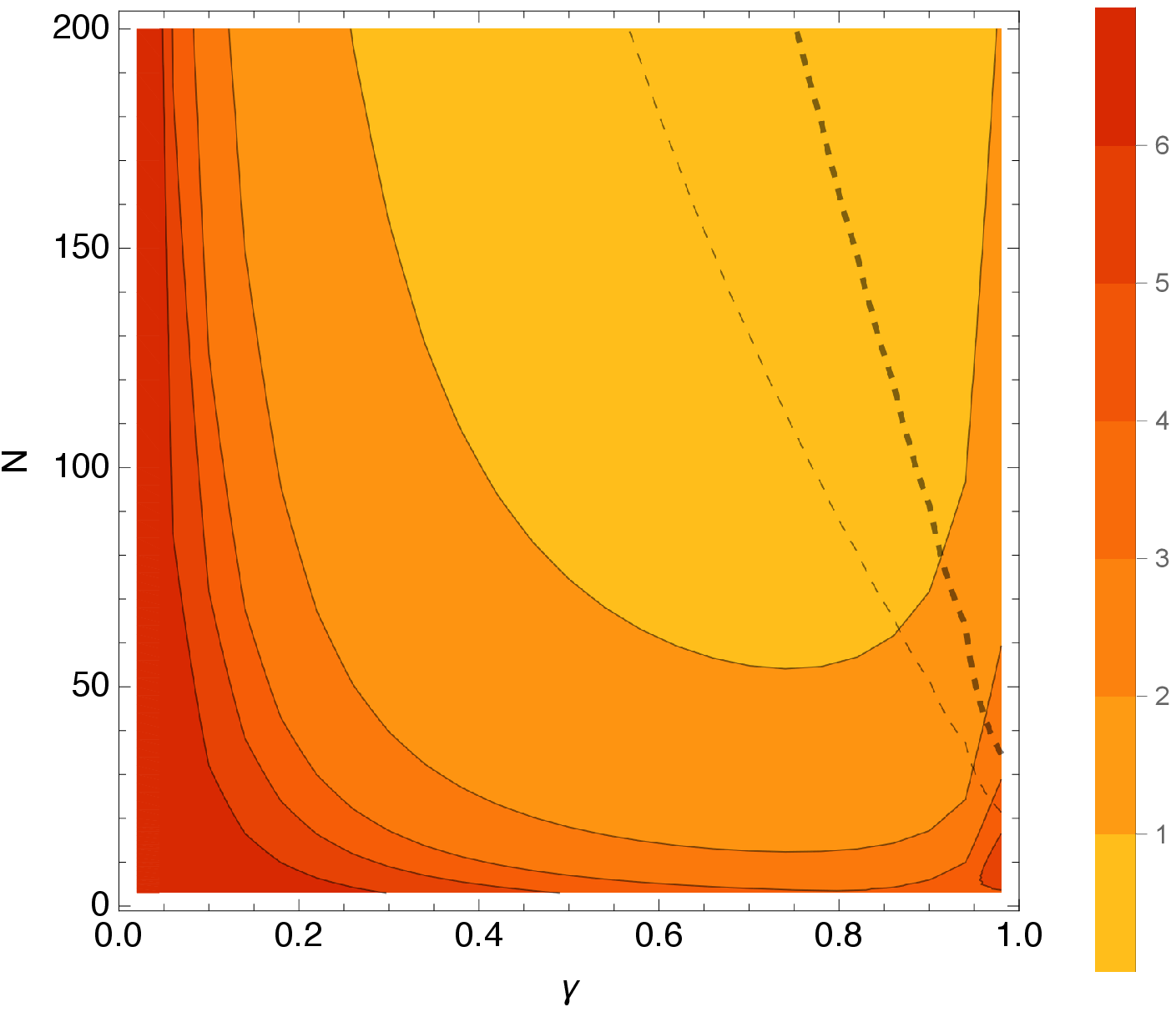}}
     \center{\includegraphics[width=.49\textwidth]{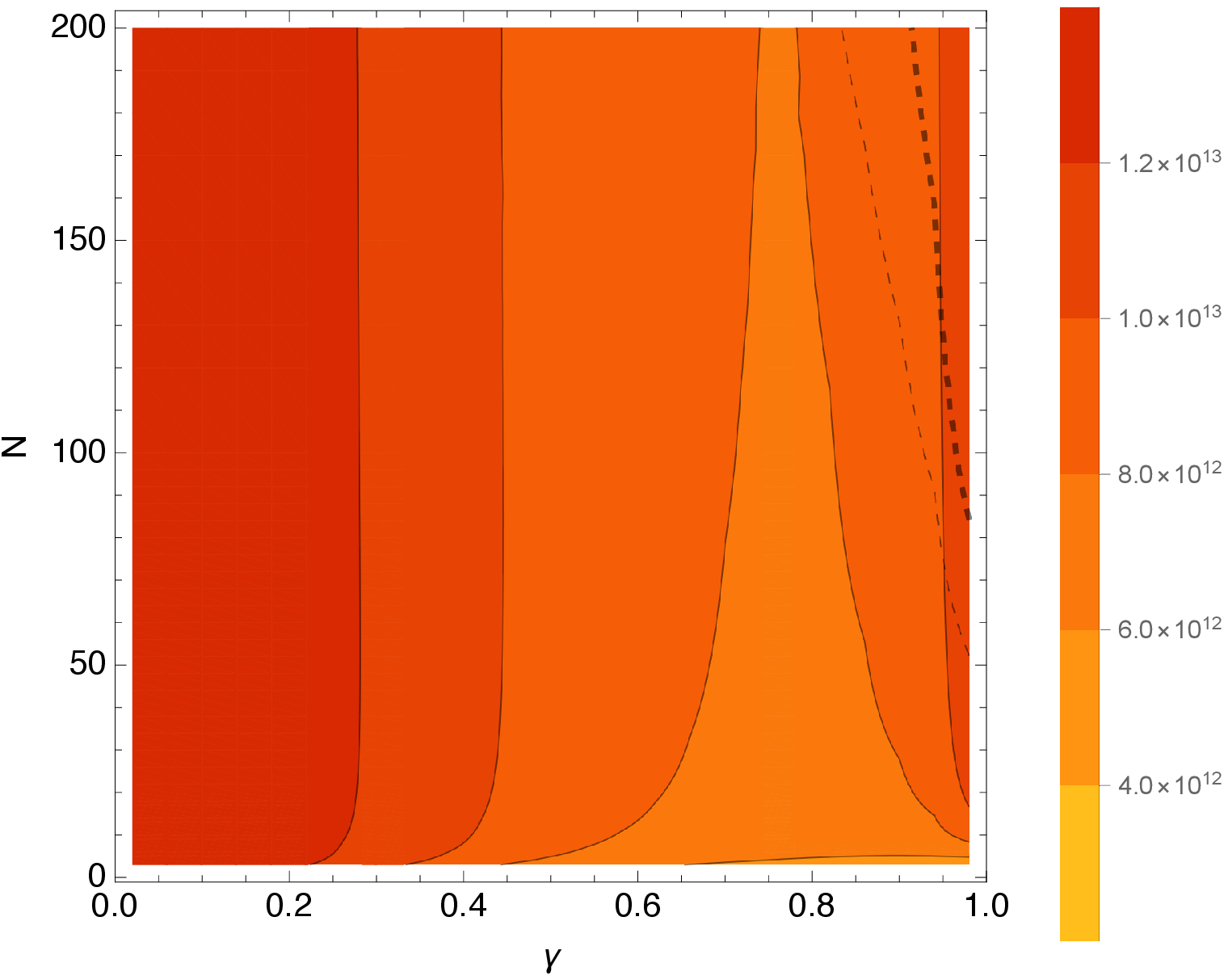} }
        \caption{%
   Most likely values of $\eta$ with $\nu=1$ and $\nu = \sqrt{N}$ (top), and $\nu=10^{-12}$ (bottom) for a red power-law power spectrum. The light and heavy dashed lines correspond to regions above which the likelihood of finding a 1-saddle is  1 in $10^{250}$ and 1 in $10^{500}$ relative to the likelihood for $\nu=-\sqrt{N}$.  }
    \label{PLPSEtaRed}
\end{figure}

Figure \ref{PLPSEtaRed} shows the range of $\eta$ produced by a red power-law power spectrum for three different scenarios: $\nu = 10^{-12}$, $\nu = 1$ and $\nu=\sqrt{N}$. The first case corresponds to the rough upper bound on $\nu$ for a ``realistic'' period of cosmological inflation, given the observed constraint on the cosmological gravitational wave background \cite{Bicep2/Planck}. The typical value of $\eta$ is very large, a consequence of the $\nu$ in the denominator of $\eta$. The $\nu=1$ and $\nu=\sqrt{N}$ cases correspond to saddles at $V \sim M_P^4$, depending on whether we constrain the typical magnitude of $V$ or the typical peak of $V$ to be Planckian, as discussed in Section~\ref{general}. These choices correspond to setting $\sigma_0 \approx M_P^4$ and $\sigma_0 \approx M_P^4/\sqrt{N}$, respectively.  In both cases the expected values of $\eta$ are not as large, and in the latter  typical values of $\eta$ can be less than unity.  

While there are values of $N$ and $\gamma$ for which we don't expect any 1-saddles with $\nu \geq 0$, the power-law spectrum can produce a full range of $\gamma$. In order to identify parameter values for which the vast majority of 1-saddles are at negative $V$ we overplot contours at which the  likelihood of a given 1-saddle occurring at a positive value of $\nu$ is (roughly) 1 in $10^{250}$ and 1 in $10^{500}$. For $\nu = \sqrt{N}$  there is a range of parameter space with $N\sim{\cal{O}}(100)$ where saddles are plentiful but $\eta$ is at the lower end of its range, including regions with $\eta<1$.

The analysis of the blue spectrum follows a similar course to the red case, albeit with a less satisfying conclusion.  The moments are
\begin{align} \label{BluePowerLawSigmas}
    \sigma_0 &= \frac{1}{(2\pi)^{N/2}}\sqrt{\frac{A k_{cut}^{N-n}}{N-n}} \quad (n < N) \, ,\\
    \sigma_1 &= \frac{1}{(2\pi)^{N/2}}\sqrt{\frac{A k_{cut}^{2+N-n}}{N-n+2}} \quad (n < N + 2) \, ,\\
    \sigma_2 &= \frac{1}{(2\pi)^{N/2}}\sqrt{\frac{A k_{cut}^{4+N-n}}{N-n+4}} \quad (n < N + 4)\, ,
\end{align}
from which we get
\begin{equation} 
    \gamma = \frac{\sqrt{(n-4-N)(n-N)}}{N - n + 2} \, ,
\end{equation}
valid only if $n < N$. An analogous substitution to that used with the red cutoff, $n \rightarrow N - \delta$ yields $\gamma = \sqrt{\delta(4+\delta)}/(\delta+2)$, an exact match to the expression found for the red cutoff and $\delta$ again measures the departure from flatness. Putting in the expressions for the moments,  
\begin{equation}
    \eta = M_P^2 \frac{\lambda}{\nu} k_{cut}^2 \sqrt{\frac{\delta}{4+\delta}} \, .
\end{equation}
In this case, the correlation function is
\begin{align}
     \xi(|\phi|) =& \frac{A}{(2\pi)^N}\int_{0}^{k_{cut}} dk k^{N-n-1}(\phi_1 k)^{1-N/2} J_{N/2-1}(\phi_1 k)         \, \\ =& \frac{A}{(2\pi)^N}2^{-N/2}k_{cut}^{N-n}\Gamma\left[\frac{N-n}{2}\right] {_1}\tilde{F}_2\left(\frac{N-n}{2};\frac{N-n}{2} + 1,\frac{N}{2};-\frac{1}{4} k_{cut}^2 \phi ^2\right). 
\end{align}

\begin{figure}[tb]
    \centerline{\includegraphics[width=.55 \linewidth]{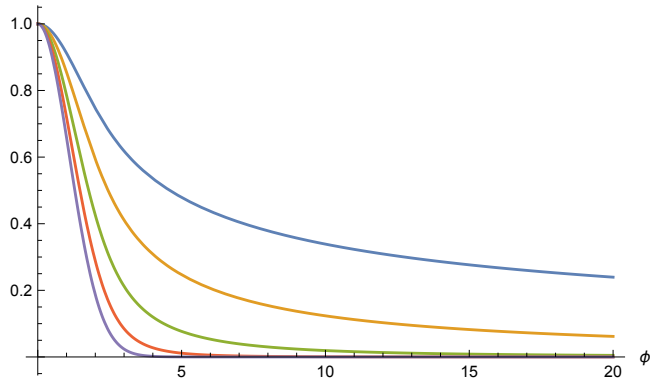}}
    \caption{%  
We plot the correlation function for the blue power-law power spectrum, for $N=50$ and (from the top) with $\delta = 0.5, 1, 2, 4$ and 10, and $k_{cut} = \sqrt{50}$. For small $\delta$ the correlation length can be arbitrarily large -- but for sufficiently large $\delta$ it tends to a fixed value. 
    }
    \label{fig:bluecorr}
\end{figure}

We can again set $k_{cut} = \sqrt{N}/M_{P}$, as the correlation function has the same scaling behaviour with $N$. However the width of the correlation function now also shows a significant dependence on $\delta$ and becomes arbitrarily large as $\delta$ approaches zero, as shown in Figure~\ref{fig:bluecorr}. Physically, this is associated with the power spectrum being increasingly weighted toward modes with very small $k$,  introducing a long-range coherence into the landscape. We could suppress this with either a second cut at small $k$, leading to results similar to the red case, or by setting larger and larger values of $k_{cut}$ as $\delta$ decreases, which would  increase $\eta$ as well. Alternatively, we can exclude scenarios with $\delta \lesssim 2$ or $\gamma \lesssim \sqrt{3}/2 \approx 0.866$.

\begin{figure}[tb]
    \center{
    \includegraphics[width=.48\textwidth]{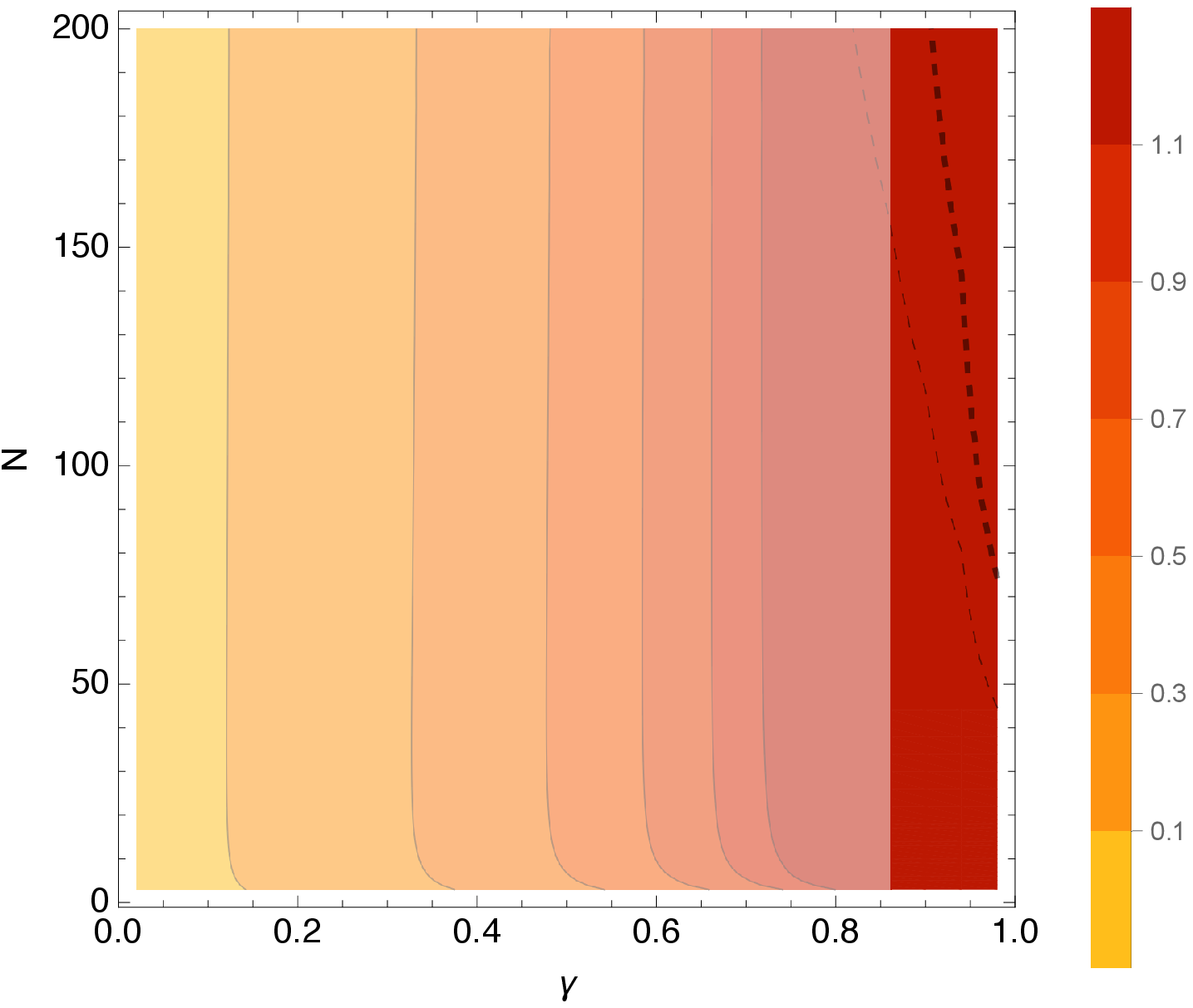} \hfill 
    \includegraphics[width=.48\textwidth]{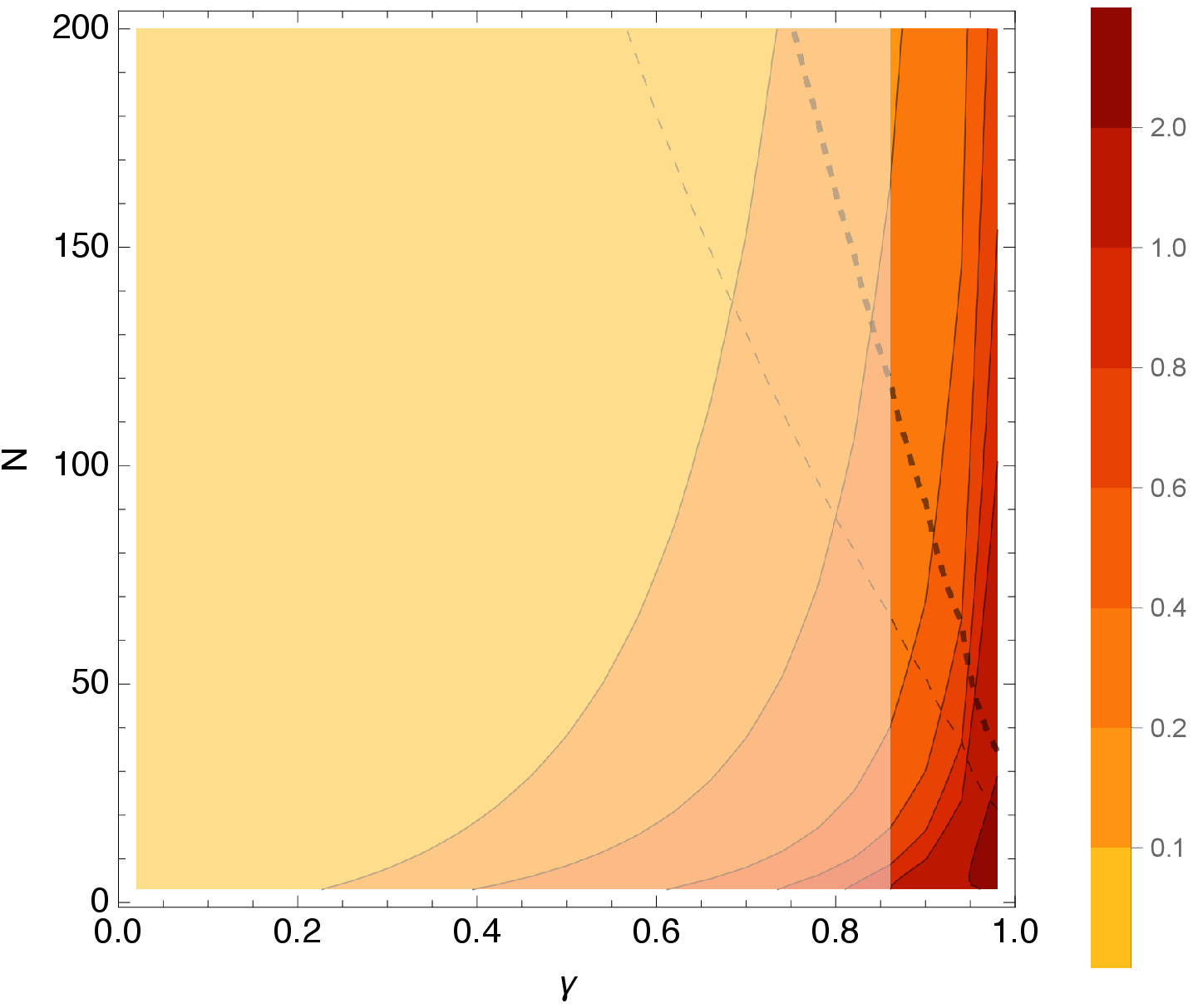}}
    
     \center{\includegraphics[width=.48\textwidth]{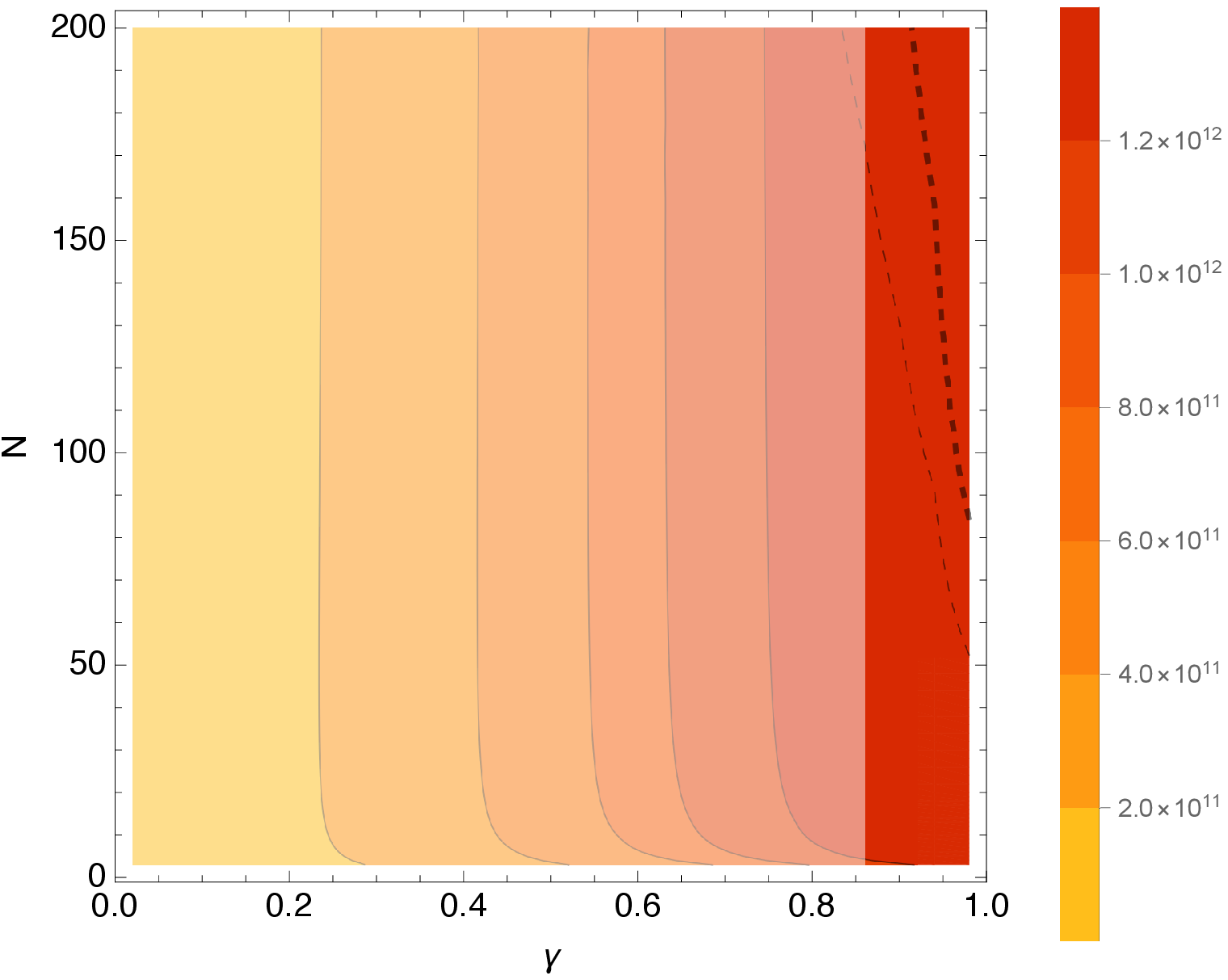}}
        \caption{Most likely values of $\eta$ with $\nu=1$ and $\nu = \sqrt{N}$ (top)  and $\nu=10^{-12}$ (bottom) for a blue power-law spectrum. The light and heavy dashed lines correspond to regions above which the likelihood of finding a 1-saddle is  1 in $10^{250}$ and 1 in $10^{500}$ relative to the likelihood for $\nu=-\sqrt{N}$; regions excluded by requiring $\delta\gtrsim 2$ are ``greyed out''; no domains with small $\eta$ and non-vanishing number of minima survive the combined cuts.}
    \label{PLPSEtaBlue}
\end{figure}
Keeping $k_{cut} = \sqrt{N}/M_{P}$, independent of $\delta$, and not introducing a second cut at small $k$, we show the most likely values of $\eta$ from a blue power-law power spectrum with the three $\nu$ values in Fig. \ref{PLPSEtaBlue}. Requiring $\gamma \gtrsim 0.866$  excludes all the regions of the parameter space with smaller $\eta$ which do not also have very small numbers of 1-saddles with $V > 0$.  

\section{Inflationary Consequences}  
\label{sec:inflation}

\begin{figure}[tb]
    \center{
    \includegraphics[width=.48\textwidth]{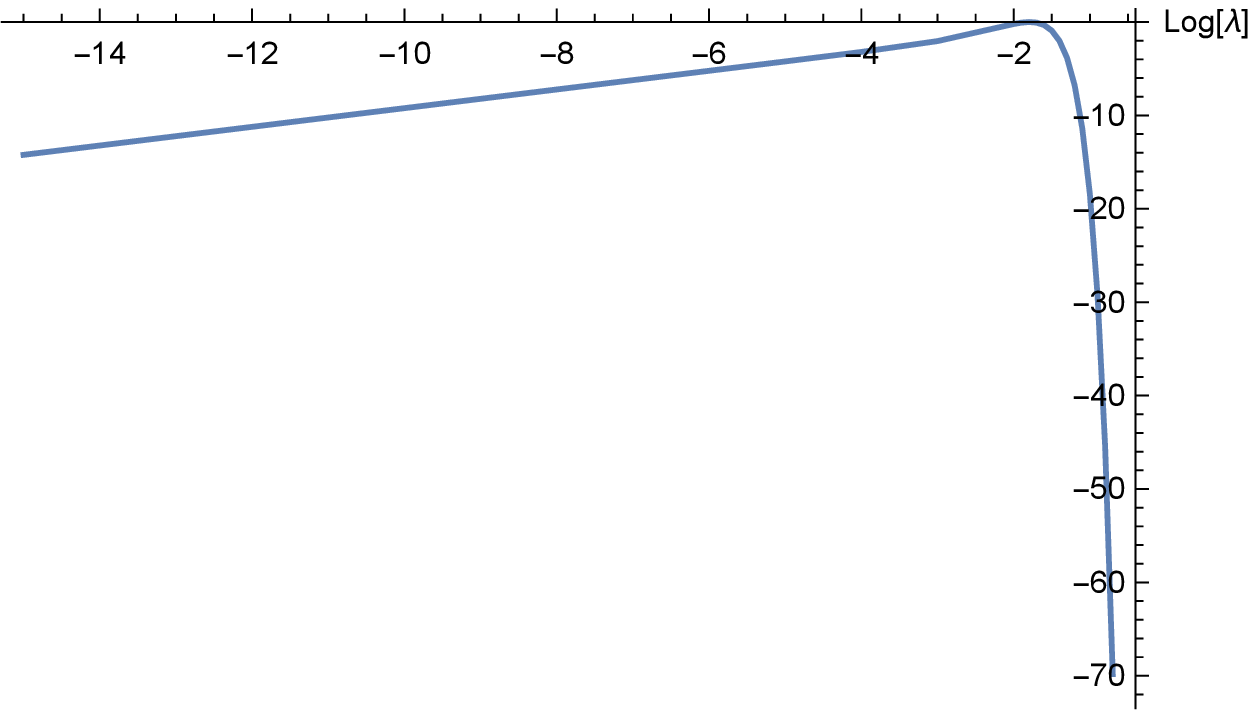} \hfill 
    \includegraphics[width=.48\textwidth]{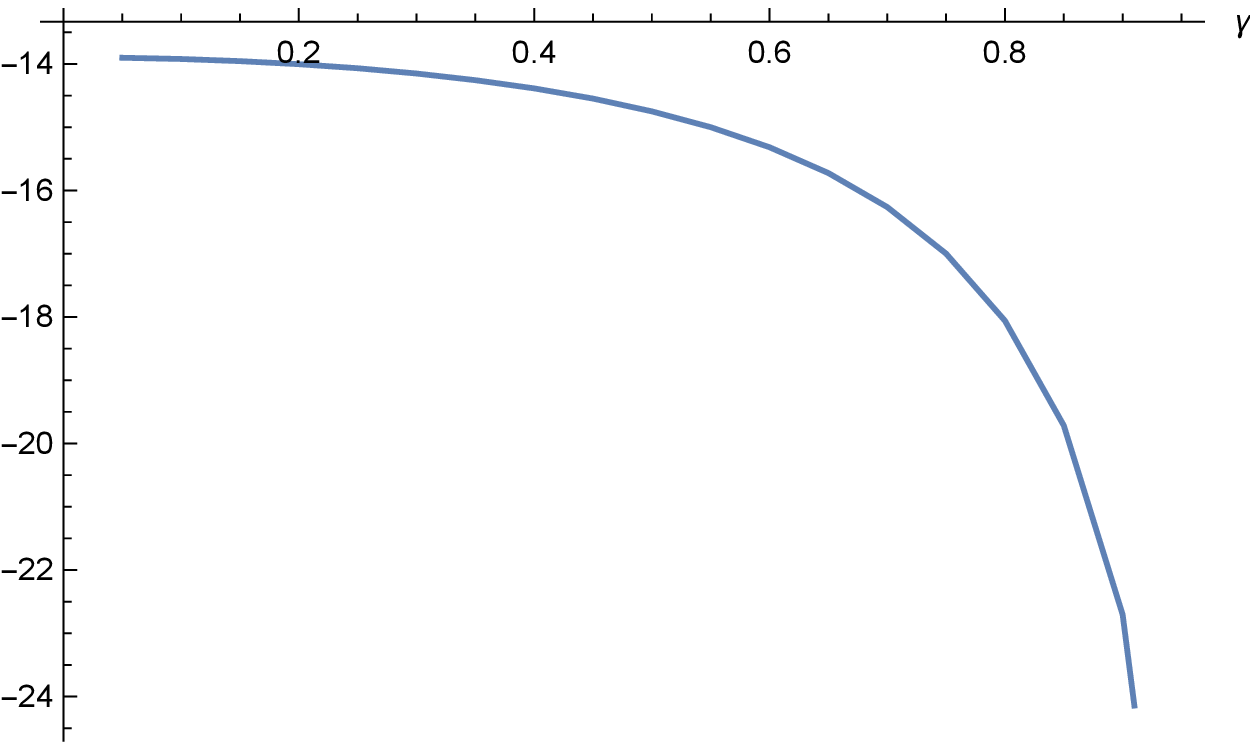}}
    \caption{%
         (Left) The likelihood of a saddle with fixed eigenvalue for $\nu = 10^{-12}, \gamma = 0.1, N = 100$. On the vertical axis is the log-likelihood with the most probable eigenvalue normalized to 0. On the horizontal axis is $\log(\lambda_{100})$, the log of the downhill eigenvalue in dimensionless units. We can see that even a very small eigenvalue of $10^{-15}$ is still only $\mathcal{O}(10)$ orders of magnitude less likely than the most probable eigenvalue. (Right) The relative log-likelihood of a 1-saddle with $\lambda = 10^{-15}$ (thus providing $\eta < 1$) compared to the most probable value of $\lambda$, as a function of $\gamma$, for $\nu = 10^{-12}, N = 100$. For small $\gamma$, these viable 1-saddles are $\approx 10^{-14}$ orders of magnitude less probable than the most probable 1-saddle, but they get less probable with increasing $\gamma$.}
    \label{PLPSeigenlog}
\end{figure}

Given the  above results we restrict attention to landscapes with a red spectrum in what follows.  From Fig. \ref{PLPSEtaRed}  we see that the most probable saddles at cosmologically-interesting values of $\nu \lesssim 10^{-12}$ do not produce viable inflation. However, inflation will be possible at rare, anomalously flat, saddles. We can quantify this  by maximising the likelihood  (Eq. \ref{Likelihood}) for specified values of $\nu$ and fixed, small values of the downhill eigenvalue, which yields an approximate probability distribution for the downward slopes of 1-saddles. This is shown in Fig. \ref{PLPSeigenlog} for $\nu = 10^{-12}, \gamma = 0.1, N = 100$. The general shape of the plot is largely independent of the specific parameter choices, but the fraction of 1-saddles which can support inflation decreases with $\gamma$. For moderate values of $\gamma$, viable 1-saddles are $\mathcal{O}(10)$ orders of magnitude less probable than typical 1-saddles. For larger $\gamma$, 1-saddles of all kinds will be rare at positive $V$ and the fraction of these with $\eta<1$ decreases rapidly when $\gamma \gtrsim 0.5$. As a consequence, the number of viable saddles at large $\gamma$ is further supressed, relative to the situation at small~$\gamma$. 

Looking at the $G$ factor in the likelihood expression shows that in this regime the likelihood is controlled by the proportionality of the likelihood to each of the individual eigenvalues. The bottom plot of Fig.~\ref{PLPSEtaRed}   suggests that in order to get $\eta < 0.01$ for a saddle which could have generated the cosmologically relevant phase of inflation, we need a $\lambda_N$ that is about $\sim 15$ orders of magnitude less probable than the peak $\lambda$ (which is the $\lambda$ used to generate this plot).  In a landscape with $\sim 10^{500}$ minima we can expect a similar number of 1-saddles, so this 15-order difference is insignificant, provided that $\gamma$ is relatively small. 
 
Of course, successful inflation requires more than a suitable saddle which is flat enough to produce a sufficient amount of inflation -- the inflaton must settle into a minimum just far enough above zero to satisfy constraints on the dark energy density.  Assessing this quantitatively would require the construction of a joint probability distribution for adjacent extrema, which is beyond the scope of this work. However, if we ask that there exists a cosmologically viable 1-saddle in the basin of attraction of a mininimum with a suitably small vacuum energy, the expected number of successful regions is 
\begin{equation}
(10^{-12} \times  N_\textrm{saddle}  \times  P(\nu > 0 | \textrm{viable 1-saddle})) \times (10^{-123} \times N_\textrm{minima} \times P(\Lambda > 0 | \textrm{minimum}) ).
 \end{equation}
The factors of $10^{-12}$ and $10^{-123}$ represent the ratios of the energy densities of the inflationary epoch and dark energy to the Planck scale, respectively. This number is much smaller than the odds of getting either a workable inflationary saddle or a viable vacuum energy on their own.  Multiplying the probabilities by these factors assumes the probability density for $\nu$ is uniform between 0 and 1. However, since the width of the full probability density function for $\nu$ is ${\cal O}((0.1-10) M_P)$ this approximation \emph{will} be correct to a few orders of magnitude, which is sufficient here. However, note that the specific inflationary trajectory will likely be modified by higher order terms in the effective potential as the field rolls away from the saddle, given that the simple inverted quadratic will quickly reach negative values.

In addition to slow-roll, topological inflation is also a possibility in the landscape. This was originally described with reference to the prototype potential \cite{Vilenkin1994}
\begin{equation}
    V(\phi) = \frac{g}{4}(\phi^2-\Delta^2)^2 \, 
\end{equation}
which supports a  classical domain wall solution, over which $\phi$ varies (spatially) between $-\phi$ and $\phi$ \cite{Vachaspati2010}. The key condition for topological inflation to occur is that the domain wall must be wider than the Hubble radius. The field is required to be continuous and therefore the domain wall cannot causally decay, so there will be a permanent ``patch'' of spacetime in which $V\sim V_0=g\Delta^4/4$ (where $V_0$ is the peak value of the potential) leading to eternal inflation at the centre of the defect.   The Hubble radius is $1/H$ or $H_0^{-1} = \sqrt{3}M_{P}/V_0^{1/2}$.

The domain wall is a static solution with the gradient energy offset against the potential energy, so the Klein-Gordon equation has the form \cite{Vachaspati2010}
\begin{equation}
    \frac{1}{2} \left(\frac{\partial \phi}{\partial x} \right)^2 = V(\phi) \, .
\end{equation}
For the prototype single-field potential above, there is an exact solution
\begin{equation}
   \phi(x) = \Delta \tanh{\left( \sqrt{\frac{g}{2}} \Delta x \right)} \, .
\end{equation}
where $x$ is the spatial coordinate. However, we can also write it in the form 
\begin{equation}
 \phi(x) = \sqrt{\frac{|V''_0|}{V_0}} \tanh{\left( \sqrt{\frac{|V''_0|}{2}} x\right)} 
\end{equation}
where $V''_0$ is
\begin{equation}
    V''_0 = \left. \frac{d^2 V}{d\phi^2} \right|_{\phi=0} \, .
\end{equation}
This can be understood as an approximate expression for a field configuration at any 1-saddle, where $V''$ is the second derivative of the potential in the downhill direction. 
 
Taking the width of the defect to be $\sqrt{2/|V''_0|}$, we can write a criterion for eternal inflation at a 1-saddle 
\begin{equation}  \label{TopologicalInflationEigenvalues}
|V''_0| \lesssim \frac{2}{3} \frac{V_0}{M_P^2}   \, .
\end{equation}
Given that eternal inflation is taking place \cite{Vilenkin:1983xq} the portion of the defect near the peak can only respond to the peak of the potential, suggesting that the approximation will be self-consistent.

The above expression is equivalent to $\eta < 2/3$. We saw in Figure \ref{PLPSEtaRed} that such values of $\eta$ are possible with the red power-law power spectrum for large enough $V$. We also saw in Figure \ref{PLPSeigenlog} that such values are possible for smaller $V$ if we go far enough into the tail of the distribution for $\lambda$. Note that the condition for topological inflation, $\eta <2/3$, is less restrictive restrictive than the observational constraint on slow-roll inflation $\eta \sim 0.01$ and therefore topological inflation will be more common than slow-roll inflation in such a landscape.

\section{Discussion}
\label{sec:discuss}

We have examined the inflationary properties of  simple Gaussian random landscape models, with either Gaussian or power-law spectra. The only scenario that is consistent with physical expectations -- that is, with non-vanishing numbers of 1-saddles and without  long-range (i.e. super-Planckian) correlations in fieldspace is the red power-law.  It is conceivable that even if the detailed  form of the landscape cannot be obtained, generic properties such as the power-spectrum might be derivable from physical considerations, so these results may potentially be used to constrain landscape scenarios.   

Separately, we have seen that the slope in the single downhill  direction leading away from a 1-saddle  decreases with $N$, relative to the uphill directions.  This behaviour reflects the deeper result that large gaps in the eigenvalue distribution of the Hessians of random landscapes are disfavoured, this is in constrast to the better-known phenomenon of eigenvalue repulsion. The eigenvalues want to be separate, but not too separate. This is a nontrivial result, insofar as it was not obvious from the outset that the one downhill eigenvalue would effectively decrease with $N$. As a consequence it appears that the $\eta$-problem of single field inflation may be softened if this scenario is assumed to be embedded in a larger landscape. That said, we did not encounter scenarios for which $\eta$ becomes arbitrarily small when $N\sim{\cal{O}}(100)$. The examination of the $\eta$-problem in  more complex scenarios, such as those where the potential is a nontrivial function of a Gaussian random function (or several functions) rather than simply a single Gaussian random field, will be a fruitful topic for further work.

For values of $V$ at which 1-saddles are plentiful, the eigenvalue distribution depends only weakly on $V$. Consequently (recalling that $\eta \propto \lambda/V$) the expected value of $\eta$ becomes larger in rough proportion to $1/V$ -- so for $V\lesssim 10^{-10} M_P^4$ the {\em typical} 1-saddle is far too steep to support inflation.\footnote{Recall that for cosmologically viable hilltop inflation \cite{Boubekeur2005} the potential is much wider than it is high; these saddles are much narrower than their typical height.} However, as shown in Figs.~\ref{PLPSEtaRed} and \ref{PLPSeigenlog}, even if $V\ll M_P^4$ a nontrivial fraction of saddles can have suitable values of $\eta$. Saddles with large positive $V$ will be rarer than those with smaller $V$, but their typical $\eta$-values will be smaller.  An interesting extension of this work would be to look more carefully at the joint distribution for the density of saddles as a function of $\eta$ and $V$, as a precursor to building a measure for the landscape. 

Anthropic ``solutions'' to the cosmological constant problem explain the present-day value of the vacuum energy by noting that if it was substantially larger than the observed value, structure formation, and thus the existence of  observers, would be suppressed. However, there is no similar  bound on an inflationary scale.\footnote{There may be limits on a gravitational wave background serving as dark radiation which is large enough to disrupt (as opposed to merely modifying) Big Bang Nucleosynthesis, but these are not strongly constraining.} For our universe, observational constraints on the gravitational wave background suggest that  $0< V \lesssim 10^{-10}M_P^4$ \cite{Bicep2/Planck}. Given that the likelihood  will not change dramatically as  $V$ changes by $10^{-10}M_P^4$ it follows that in a multiverse with a nontrivial number of 1-saddles at positive $V$, the ``typical'' inflationary 1-saddle will have $V>10^{-10}M_P^4$ and this yields a stochastic  gravitational wave background inconsistent with our Universe. To whatever extent that this constitutes a prediction of this scenario, it is arguably disfavoured -- but not excluded -- by observations.  

We have  focused on the properties of the landscape potential. However, the hypothesis underlying this work is that the landscape supports a cosmological multiverse, a vast number of causally disconnected pockets of spacetime. While we cannot make direct observational tests of a multiverse hypotheses, this analysis  demonstrates that it may be possible  to place non-trivial quantitative constraints on specific landscape proposals by demonstrating that they cannot be expected to provide  a single viable inflationary 1-saddle. These considerations immediately eliminated random landscapes with a  Gaussian power spectrum, and made those with a blue power-law spectrum appear inconsistent without the introduction of a tuning. Furthermore, as noted above, even the surviving scenario with a red power-spectrum may ``predict'' a gravitational wave background at odds with present cosmological observations. The nontrivial nature of these results illuminates ways in which it is possible to draw quantitative inferences about specific multiverse proposals. 
   
 \acknowledgments We acknowledge useful discussions with  Mateja Gosenca,  Ali Masoumi,  Mark Mueller and Jens Niemeyer. This work was supported in part by the Foundational Questions Institute (FQXi)  and the Silicon Valley Community Fund.

\end{document}